\documentclass[11pt]{JHEP3}

\usepackage{subfigure, axodraw, amsmath}
\usepackage{cite}

\newcommand{\mhvbar}{$\overline{\text{MHV}}$}

\newcommand{\tr}{{\mathrm{tr}}}
\renewcommand{\vec}[1]{{\mathbf{#1}}}
\newcommand\mdot{\!\cdot\!}
\newcommand\gF{\mathcal{F}}
\newcommand\gA{\mathcal{A}}
\newcommand\gB{\mathcal{B}}
\newcommand\pD{\mathcal{D}}

\newcommand\mpp{\text{$-$$+$$+$}}
\newcommand\ppm{\text{$+$$+$$-$}}

\newcommand\fourplus{\text{$+$$+$$+$$+$}}

\newcommand\ie{\textit{i.e.}}
\newcommand\eg{\textit{e.g.}}
\newcommand\cf{\textit{cf.}}

\newcommand{\aka}{{a.k.a.}\ }

\title{%
S-Matrix
Equivalence Theorem Evasion and Dimensional Regularisation with the
Canonical MHV Lagrangian
}

\author{%
James~H.~Ettle${}^*$,
Chih-Hao~Fu${}^\dag$,
Jonathan~P.~Fudger${}^*$,
Paul~R.~W.~Mansfield${}^\dag$ and
Tim~R.~Morris${}^*$\\
\begin{tabbing}
${}^*$\;\=%
School of Physics and Astronomy,  University of Southampton\\
\>Highfield, Southampton, SO17 1BJ, U.K.\\ 
${}^\dag$\>%
Department of Mathematical Sciences, University of Durham\\
\>South Road, Durham, DH1 3LE, U.K.\\ 
E-mails:\;\=
\email{jhe@phys.soton.ac.uk}, \email{chih-hao.fu@durham.ac.uk},\\
\>\email{j.p.fudger@phys.soton.ac.uk}, \email{P.R.W.Mansfield@durham.ac.uk},\\
\>\email{T.R.Morris@soton.ac.uk}
\end{tabbing}
}


\preprint{SHEP 07-09}

\abstract{%
  We demonstrate that the canonical change of variables that yields the MHV
  lagrangian, also provides contributions to scattering amplitudes that evade
  the equivalence theorem. This `ET evasion' in
  particular provides the tree-level $(\mpp)$ amplitude, which is
  non-vanishing off shell, or on shell with complex momenta or in $(2,2)$
  signature, and is missing from the MHV (\aka CSW) rules. At one loop there
  are ET-evading diagrammatic contributions to the amplitudes with all
  positive helicities.  We supply the necessary regularisation in order to
  define these contributions (and quantum MHV methods in general) by
  starting from the light-cone Yang-Mills lagrangian in $D$ dimensions and
  making a canonical change of variables for all $D-2$ transverse degrees of
  freedom of the gauge field.  In this way, we obtain dimensionally
  regularised three- and four-point MHV amplitudes. Returning to the
  one-loop $(\fourplus)$ amplitude, we demonstrate that its quadruple cut
  coincides with the known result, and show how the original light-cone
  Yang-Mills contributions can in fact be algebraically recovered from the
  ET-evading contributions. We conclude that the canonical MHV
  lagrangian, supplemented with the extra terms brought to correlation
  functions by the non-linear field transformation,
  provide contributions which are
  just a rearrangement of those from light-cone Yang-Mills and thus coincide
  with them both on and off shell.
}

\maketitle

\begin{document}

\section{Introduction}
\label{sec:intro}

Much recent activity in perturbative QCD has revolved around a variety of
new techniques for calculating Yang-Mills scattering amplitudes that have
considerable computational advantages over the traditional Feynman graph
approach.

Parke and Taylor obtained very compact expressions for the MHV (and \mhvbar)
amplitudes \cite{Parke:1986gb, Berends:1988zn, Parke:1989vn}, which were
found to have a simple geometrical interpretation in twistor space
\cite{Witten:2003nn}. Taking inspiration from twistor-string theory,
Cachazo, Svr\v cek and Witten formulated the MHV rules (or CSW rules)
\cite{Cachazo:2004kj} for the construction of tree-level scattering
amplitudes.  These rules treat MHV amplitudes as vertices, and continue them
off-shell by a particular prescription. The vertices are joined by scalar
propagators. These rules have since seen extensive application (\eg\ in
\cite{Georgiou:2004wu, Wu:2004fb, Bena:2004ry, Wu:2004jx, Kosower:2004yz,
  Georgiou:2004by}) and have been proved a number of ways (such as
\cite{Britto:2005fq} and \cite{Risager:2005vk}). Britto, Cachazo, Feng and
Witten \cite{Britto:2004ap, Britto:2005fq, Britto:2005dg} discovered
recursion relations between on-shell tree-level scattering amplitudes (the
so-called BCFW approach) that ultimately reduce everything to sums of
products of the MHV and \mhvbar\ amplitudes.

These methods have not been as forthcoming for quantum-level computations.
Nevertheless, considerable ingenuity has yielded several results for
perturbative Yang-Mills at one-loop (including amplitudes with any number of
gluons where at most two gluons have a different helicity to the others
\cite{Bern:1993qk, Mahlon:1993si, Bern:2005hs, Berger:2006vq}, and certain
configurations for up seven gluons \cite{ Britto:2004nj, Bern:2005cq }) by
application of unitarity, generalised unitarity and cut construction
\cite{Bern:1994cg, Brandhuber:2005jw}, on-shell recursion
relations/bootstrapping \cite{Bern:2005hs, Bern:2005ji, Bern:2005hh,
  Bern:2005cq, Berger:2006ci}, the holomorphic anomaly \cite{Cachazo:2004zb,
  Cachazo:2004by, Cachazo:2004dr, Britto:2004nj}, using the CSW rules within
loops, evaluating the diagrams with the Feynman tree theorem
\cite{Brandhuber:2004yw, Brandhuber:2005kd}, and hybrid techniques
extracting the rational pieces from the Feynman diagrams \cite{Xiao:2006vr}.
Being entirely cut-constructable, ${\cal N}=1$ and ${\cal N}=4$ SYM are
particularly amenable to generalised unitarity approaches, examples of which
can be found in \cite{Britto:2004nc, Bidder:2005ri, Buchbinder:2005wp,
  Risager:2005ke, Bedford:2004nh, Bern:2004bt, Quigley:2004pw,
  Bedford:2004py, Brandhuber:2004yw}. Indeed, Yang-Mills amplitudes can be
written as a linear combination of ${\cal N}=1$ and ${\cal N}=4$ pieces,
leaving a scalar loop \cite{Bedford:2004nh, Brandhuber:2005jw}. The latter
may be evaluated using the methods discussed above, including novel
approaches to generalised unitarity outside four dimensions that recover the
rational parts \cite{Brandhuber:2005jw}. Nevertheless, a systematic,
tractable approach to calculating \emph{complete} one-loop amplitudes with
MHV techniques is yet to be found.

The CSW approach applied at the tree and quantum level, is highly suggestive
of a quantum field theory, but up until recently it was not formulated in
the lagrangian framework which would allow us to use the full machinery of
quantum field theory; instead it has been supported by conjecture,
demonstration and varying degrees of proof. Progress of particular interest
to the work herein was presented in refs. \cite{Gorsky:2005sf} and
\cite{Mansfield:2005yd}, where a \emph{canonical} transformation of the
field variables was described that mapped the self-dual sector of Yang-Mills
on to [the kinetic term of] a free field theory. Doing so resulted in a
lagrangian with an infinite set of terms, each forming an MHV vertex. These
vertices would be joined by scalar propagators following the CSW
prescription; that scattering amplitudes calculated with the new vertices
matched the traditional results at tree-level followed providing the
equivalence theorem \cite{Bergere:1975tr, Itzykson:1980rh} was satisfied.
This transformation was made explicit in \cite{Ettle:2006bw} where it was
demonstrated that the terms were MHV vertices (for up to five gluons).

In the meantime, other developments invoking similar techniques driven by
field transformation have arisen. Feng and Huang \cite{Feng:2006yy} obtained
an MHV lagrangian for ${\cal N}=4$ Super-Yang-Mills in light-cone superspace
by a change of the superfield variables. Brandhuber, Spence and Travaglini
\cite{Brandhuber:2006bf} consider just a holomorphic change of variables,
arguing that the one-loop all-$+$ amplitude then arises from the
transformation's jacobian. Boels, Mason and Skinner \cite{Mason:2005kn,
Boels:2006ir, Boels:2007qn, Boels:2007gv} recover MHV diagrams using a
twistor action with which they show that the formalism arises from a gauge
transformation on the larger twistor space (as opposed to a non-linear field
redefinition).

In this paper, we develop the ideas of \cite{Mansfield:2005yd,
Ettle:2006bw}. We have already noted there that the treatment of the
canonical MHV lagrangian at the quantum level required proper regulation,
and that certain amplitudes known not to vanish could not be constructed
from its vertices. Here, we demonstrate that the missing amplitudes arise
naturally from application of the standard procedure of LSZ reduction
\cite{Itzykson:1980rh} to the theory's correlation functions, in this case
leaving behind evasions of the equivalence theorem. We incorporate
dimensional regularisation by applying the canonical transformation to all
the transverse degrees of freedom of light-cone Yang-Mills (henceforth LCYM)
in $D=4-2\epsilon$ dimensions. This is then used to investigate the
four-gluon all-$+$ amplitude.

In section \ref{sec:cmhv-4d-review}, we review the construction of
the canonical MHV lagrangian in four dimensions. We show how the
construction evades the equivalence theorem in section
\ref{sec:treelevel-etviol}, and use this to recover the $(\mpp)$
amplitude. This is non-vanishing on shell with complex momenta or
in $(2,2)$ signature but we see that we recover the three-point
\mhvbar\ amplitude even off shell with the appropriate CSW
prescription.

Section \ref{sec:dimreg-mhvl} derives the canonical MHV lagrangian
in $D$ dimensions. We obtain the series coefficients that specify
the change of variables and the appropriate generalisations of the
three- and four-point MHV amplitudes. We note that the series
coefficients can be written in terms of the three-point \mhvbar\
vertex that was eliminated in the canonical transformation,
emphasising the generality of this procedure. This step helps
clarify how the missing amplitudes are recovered, which we do in
the next section.

We then analyse the one-loop $(\fourplus)$ amplitude. We show that at the
algebraic level, \ie\ before taking on-shell limits or performing the loop
integral, that the ET-evading contributions sum up to precisely the
missing amplitudes that LCYM would have provided. We concentrate on the LCYM
box contribution, which we uncover by tracing the momentum routing through
the three-point \mhvbar\ vertices. Although we concentrate on this one
topology it is clear that all the LCYM contributions are recovered in this
way.

We also analyse cuts of the ET-evading contributions,
demonstrating that the quadruple cut of the propagators coincides
with that of the known one-loop $(\fourplus)$ amplitude. This
serves also to show that our off shell $D$ dimensional
generalisations of the spinor bracket techniques can be used
efficiently to compute these amplitudes.

Finally in section \ref{sec:discussion} we draw all these strands
together, make our conclusions, and indicate directions for future
research.

\section{A Review of the 4D Canonical MHV Lagrangian}
\label{sec:cmhv-4d-review}

In this section, we will briefly review the canonical MHV
lagrangian in four dimensions. This will also serve to explain the
conventions we use for this paper.

\subsection{Light-cone co-ordinates}
\label{ssec:lightcone-4d}

The construction of the canonical MHV lagrangian discussed in
\cite{Mansfield:2005yd} begins with LCYM theory
\cite{Leibbrandt:1983pj} in a co-ordinate system defined by
\begin{equation}
       x^0 = \tfrac1{\sqrt 2}(t-x^3), \quad
x^{\bar 0} = \tfrac1{\sqrt 2}(t+x^3), \quad
         z = \tfrac1{\sqrt 2}(x^1+ix^2), \quad
    \bar z = \tfrac1{\sqrt 2}(x^1-ix^2).
\label{eq:lc-4d-coords}
\end{equation}
Note here the presence of the $1/\sqrt{2}$ factors that preserve
the normalisation of the volume form. It useful to employ a
compact notation for the components of 1-forms in these
co-ordinates, for which we write $(p_0, p_{\bar 0}, p_z, p_{\bar
z}) \equiv (\check{p}, \hat p, p, \bar p)$; for momenta labelled
by a number, we write that number with a decoration, for example
the $n^{\rm th}$ external momentum has components $(\check{n},
\hat n, \tilde n, \bar n)$. In these co-ordinates and with this
notation, the Lorentz invariant reads
\begin{equation}
A \cdot B = \check{A}\,\hat B + \hat A\,\check{B} - A \bar B - \bar A B.
\label{eq:lc-4d-invariant}
\end{equation}
The following bilinears are also defined:
\begin{equation}
\label{eq:lc-4d-bilinears}
  (1\:2) := \hat 1 \tilde 2 - \hat 2 \tilde 1, \quad
\{1\:2\} := \hat 1 \bar 2 - \hat 2 \bar 1.
\end{equation}
In four dimensions, we can express these bilinears in terms of the
conventional angle and square brackets often found in the
literature. We will not make much use of the spinor formalism in
this paper, but it is nevertheless illuminating to consider this
relationship. Begin by noting that for a 4-vector $p$, its
bispinor representation is
\[
(p_{\alpha\dot\alpha}) = \sqrt{2} \begin{pmatrix}
    \check p    &  -p  \\
    -\bar p & \hat p
\end{pmatrix},
\]
and that for null $p$ this factors into $p_{\alpha\dot\alpha} =
\lambda_\alpha \tilde\lambda_{\dot\alpha}$ where we can choose
\begin{equation}
\label{eq:lc-4d-spinors}
\lambda = 2^{1/4} \begin{pmatrix} -p/\sqrt{\hat p} \\ \sqrt{\hat p}
    \end{pmatrix}
\quad\text{and}\quad
\tilde\lambda = 2^{1/4} \begin{pmatrix} -\bar p/\sqrt{\hat p} \\ \sqrt{\hat p}
    \end{pmatrix}.
\end{equation}
Hence the spinor brackets can be expressed as
\begin{equation}
\label{eq:lc-4d-spinorbrackets}
\langle 1\:2 \rangle := \epsilon^{\alpha\beta}
    \lambda_{1\alpha} \lambda_{2\beta}
    = \sqrt{2} \frac{(1\:2)}{\sqrt{\hat 1 \hat 2}}
\quad\text{and}\quad [ 1\:2 ] := \epsilon^{\dot\alpha\dot\beta}
    \lambda_{1\dot\alpha} \lambda_{2\dot\beta}
    = \sqrt{2} \frac{\{1\:2\}}{\sqrt{\hat 1 \hat 2}}.
\end{equation}
Observe that $\lambda$ and $\tilde\lambda$ shown in \eqref{eq:lc-4d-spinors}
are also defined for non-null $p$. In this case, their product corresponds to
the null vector
\begin{equation}
\lambda\tilde\lambda = p + a \mu,
\label{eq:csw-nullvector}
\end{equation}
where $\mu = \eta \tilde\eta = (1,0,0,1)/\sqrt{2}$ in Minkowski
co-ordinates is the null vector normal to surfaces of constant
$x^0$, and $a$ is a coefficient which is unimportant here. By
contracting both sides of \eqref{eq:csw-nullvector} with
$\tilde\eta$, we see that $\lambda$ satisfies the CSW prescription
\cite{Cachazo:2004kj} for continuing spinor momenta off the mass shell.

\subsection{The field transformation}
\label{ssec:lightcone-ym-4d}

The Yang-Mills action can be written as
\[
S=\frac 1{2g^2} \int d^4x\:\tr \gF^{\mu\nu}\gF_{\mu\nu},
\label{eq:ym-4d-action}
\]
where $d^4x$ is the Minkowski volume element and we define the
field-strength tensor $\gF$ by
\begin{equation}
\gF_{\mu\nu}=[D_\mu, D_\nu], \quad
D_\mu = \partial_\mu + \gA_\mu, \quad
\gA_\mu = -\frac{ig}{\sqrt 2}A^a_\mu T^a.
\label{eq:ym-defs}
\end{equation}
Our internal group generators are normalised according to the
convention of \cite{Dixon:1996wi} as
\begin{equation}
[T^a,T^b] = i\sqrt 2 f^{abc} T^c, \quad
\tr(T^a T^b) = \delta^{ab}.
\label{eq:ym-generators}
\end{equation}

Quantisation takes place on surfaces $\Sigma$ of constant $x^0$ with normal
$\mu$ defined as above. We choose the axial gauge condition $\mu~\cdot~\gA =
\hat\gA = 0$, for which the Faddeev-Popov ghosts are completely decoupled. The
lagrangian density is then quadratic in the $\check\gA$ field, and we integrate
it out of the partition function to obtain the light-cone action
\begin{equation}
\label{eq:ym-lc4d-action}
S = \frac 4{g^2} \int dx^0\:(L^{-+}+L^{\mpp}+L^{--+}+L^{'--++}),
\end{equation}
where
\begin{align}
\label{eq:ym-4d-mp}
L^{-+} &= \phantom{-}{\rm tr}\int_\Sigma d^3{\bf x}\:
{\bar \gA}(\check\partial\hat\partial-
\partial\bar\partial)\gA\\
\label{eq:ym-4d-mpp}
{L}^{-++}&=-{\rm tr}\int_\Sigma d^3{\bf x}\:
({\bar\partial}{\hat\partial}^{-1} {  \gA})\:
[{  \gA},{\hat\partial} {\bar\gA}]\\
\label{eq:ym-4d-mmp}
{L}^{--+}&=-{\rm tr}\int_\Sigma d^3{\bf x}\: [{\bar
A},{\hat\partial} {  \gA}]\:
({  \partial}{\hat\partial}^{-1} {\bar \gA})\\
\label{eq:ym-4d-mmpp}
{L}^{'--++}&=-{\rm tr}\int_\Sigma d^3{\bf x}\: [{\bar \gA
},{\hat\partial} { \gA }]\:{\hat\partial}^{-2}\: [{ \gA
},{\hat\partial} {\bar \gA }].
\end{align}

\begin{figure}[h]
  \centering \subfigure{
    \begin{picture}(50,80)
      \Line(0,0)(25,20) \Line(50,0)(25,20)
      \Line(25,20)(25,60)
      \Line(0,80)(25,60) \Line(50,80)(25,60)
      \Text(32,25)[cc]{$+$} \Text(32,55)[cc]{$-$}
      \Text(0,8)[cc]{$-$} \Text(50,8)[cc]{$+$}
      \Text(0,72)[cc]{$+$} \Text(50,72)[cc]{$+$}
    \end{picture}
    \qquad
    \begin{picture}(80,50)
      \Line(0,0)(20,25) \Line(0,50)(20,25)
      \Line(20,25)(60,25)
      \Line(80,50)(60,25) \Line(80,0)(60,25)
      \Text(0,8)[cc]{$-$} \Text(0,42)[cc]{$+$}
      \Text(80,8)[cc]{$+$} \Text(80,42)[cc]{$+$}
      \Text(27,18)[cc]{$+$}
      \Text(53,18)[cc]{$-$}
    \end{picture}
    \label{fig:feyn-lc-mppp-a}
  }\qquad\qquad \subfigure{
    \begin{picture}(80,80)
      \Line(0,0)(25,20) \Line(55,20)(25,20)
      \Line(55,20)(80,0) \Line(55,20)(80,40) \Text(80,32)[cc]{$+$}
      \Line(25,20)(25,60)
      \Line(0,80)(25,60) \Line(50,80)(25,60)
      \Text(18,25)[cc]{$+$} \Text(18,55)[cc]{$-$}
      \Text(0,8)[cc]{$-$} \Text(80,8)[cc]{$+$}
      \Text(0,72)[cc]{$+$} \Text(50,72)[cc]{$+$}
      \Text(32,12)[cc]{$+$} \Text(50,12)[cc]{$-$}
    \end{picture}
    \label{fig:feyn-lc-mppp-b}
  }
  \caption{Examples of possible tree-level diagrams that can be constructed
  from the Chalmers-Siegel truncation of LCYM.}
  \label{fig:feyn-lc-mppp}
\end{figure}
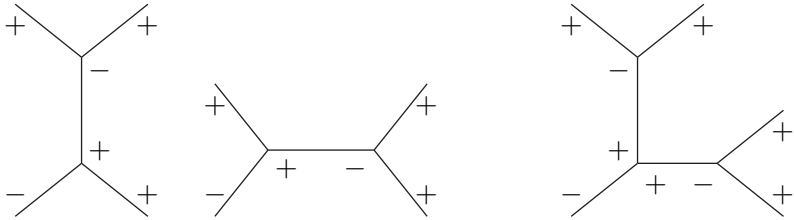

The unwanted $L^{-++}$ is removed by defining a new Lie algebra
valued field $\gB$ such that
\begin{equation}
\label{eq:transform-4d}
L^{-+}[\gA] + L^{-++}[\gA] = L^{-+}[\gB].
\end{equation}
It may seem rather perverse to absorb the three-point $\mpp$ vertex
into the kinetic term, as in \eqref{eq:transform-4d}. It makes
sense once one recognises that the left hand side is the self-dual
sector of Yang-Mills \cite{Chalmers:1996rq} which at tree level
gives only diagrams with one negative helicity and all the rest
positive, as illustrated in fig.~\ref{fig:feyn-lc-mppp}. Such
terms are well known to vanish on shell \cite{Dixon:1996wi}, so at
least in four dimensions, and at tree level and on shell, we are
simply making explicit what is already a free theory. We will
reconsider this point in more generality in the conclusions.

The transformation is performed entirely on the
quantisation surface $\Sigma$, and so all the fields have the same $x^0$
dependence, which we suppress. By restricting $\gA$ to be a
function of $\gB$ alone and requiring the transformation to be
canonical, we find that $\bar\gA$ must be a function of $\gB$ and
$\bar\gB$, and contain only one power of the latter.
Now working in momentum space on $\Sigma$, we can write the following series
solutions:
\begin{align}
\label{eq:cmhv-4d-A-series}
\gA_1 &= \sum_{n=2}^{\infty} \int_{2\cdots n}
    \Upsilon(1\cdots n) \:
    \gB_{\bar 2} \cdots \gB_{\bar n} \: (2\pi)^3
    \delta({\textstyle \sum_{i=1}^n \vec p_i}), \\
\label{eq:cmhv-4d-Abar-series}
\hat 1 \bar\gA_{\bar1} &=
\sum_{m=2}^\infty\sum_{s=2}^m\int_{2\cdots m}{\hat
s}\,\Xi^{s-1}(\bar 12\cdots m)\:\gB_{\bar2}\cdots\bar\gB_{\bar s}
\cdots \gB_{\bar m} \: (2\pi)^3 \delta({\textstyle \sum_{i=1}^m
\vec p_i}),
\end{align}
where $\Upsilon(12)= \Xi^1(12) = 1$. Note that the numbered
subscripts above are momentum arguments, and the bar implies
negation: $\gB_{\bar i} := \gB(-\vec p_i)$. The integral
short-hand here is defined by
\[
\int_{1\cdots n} \equiv \prod_{k=1}^n
    \frac1{(2\pi)^{3}}
    \int d\hat k\,dk\,d\bar k.
\]
The requirement that the transformation is canonical and
\eqref{eq:transform-4d} imply the following recursion relations
for $\Upsilon$ and $\Xi$:
\begin{align}
\Upsilon(1\cdots n) &= \frac i{\omega_1+\cdots+\omega_n}
    \sum^{n-1}_{j=2} (\zeta_{j+1,n}-\zeta_{2,j})
        \Upsilon(- ,2,\cdots,j) \Upsilon(- ,j+1,\cdots,n),
    \label{eq:Upsilon-4d-rr}\\
\begin{split}
\Xi^l(1\cdots n) &=
    - \sum_{r=\max(2,4-l)}^{n+1-l} \sum_{m=\max(r,3)}^{r+l-1}
        \Upsilon(-,n-r+3,\cdots,m-r+1) \\
    & \hphantom{=-\sum_{r=\max(2,4-l)}^{n+1-l}\sum_{m=\max(r,3)}^{r+l-1}}
        \times \Xi^{l+r-m}(-,m-r+2,\cdots,n-r+2).
    \label{eq:Xi-4d-rr}
\end{split}
\end{align}
Here, and hereafter, the arguments labelled ``$-$'' are minus the sum of the
remaining arguments (as follows from momentum conservation),
$\zeta_{j,k} := \zeta(\sum_{i=j}^k p_i)$ where $\zeta(p) := \bar
p/\hat p$, $\omega_j = {\tilde j \bar j}/{\hat j}$, and momentum
indices must be interpreted cyclically. These can be solved to
give\footnote{For $\Upsilon(123)$, in the numerator only the $\hat
1$ is retained.}
\begin{align}
\label{eq:Upsilon-4d}
\Upsilon(1\cdots n) &= i^n \frac{\hat 1 \hat 3 \cdots \widehat{n-1}}
    {(2\:3)\cdots(n-1,n)},\\
\Xi(1\cdots n) &= -\frac{\hat s}{\hat 1} \Upsilon(1\cdots n).
\label{eq:Xi-4d}
\end{align}

Putting all this together, it is not difficult to see that the
remaining pieces of the lagrangian \eqref{eq:ym-4d-mmp},
\eqref{eq:ym-4d-mmpp} are transformed into an infinite series of
terms, labelled by helicity, each with two factors of $\bar\gB$ such that
\begin{equation}
\label{eq:cmhvl-4d}
L = L^{-+}[\gB] + L^{--+}[\gB] + L^{--++}[\gB] + L^{--+++}[\gB] + \cdots.
\end{equation}
Explicit substitution shows that the vertices,
\begin{equation}
\label{eq:V}
\frac 12 \sum_{s=2}^n \int_{1\cdots n}\!\!\!\!
        V^s(1\cdots n) \: \tr [
            \bar\gB(-\vec p_1) \gB(-\vec p_2)
            \cdots \bar\gB(-\vec p_s) \cdots
            \gB(-\vec p_n)]\: (2\pi)^3\delta({\textstyle \sum_{i=1}^n \vec
            p_i}),
\end{equation}
contained therein are, indeed, the MHV amplitudes.

At this point it is important to address the normalisation of the
gauge fields. So far we have been working with a non-canonical
normalisation  from \eqref{eq:ym-lc4d-action}, \eqref{eq:ym-4d-mp}
and \eqref{eq:cmhvl-4d} that absorbs the coupling constant and
group generators; one upshot of this is that the tree-level
expression for a $\gB$ propagator is
\begin{equation}
\label{eq:gBgB-prop-4d}
\langle \gB \bar\gB \rangle = -\frac{ig^2}{2p^2}.
\end{equation}
Upon substitution for fields with the canonical normalisation
using \eqref{eq:ym-defs}, \ie\ using the same transformation also
for $\gB$ and $\bar\gB$, one finds of course the $B(\bar B)$
propagator $i/p^2$, as expected. Particular care must therefore be
taken to use the appropriate propagator \eqref{eq:gBgB-prop-4d}
and external polarisation vectors in conjunction with the series
coefficients $\Upsilon$ and $\Xi$ as defined in
\eqref{eq:Upsilon-4d} and \eqref{eq:Xi-4d}. Alternatively, one can
work (as we will do in the computations to follow) entirely with
canonical normalisation by making the replacements
\begin{align}
V^s(1\cdots n) &\rightarrow
    \frac4{g^2} \left(-\frac{ig}{\sqrt 2}\right)^{n}
    V^s(1\cdots n), \\
\Upsilon(1\cdots n) &\rightarrow
    \left(-\frac{ig}{\sqrt 2}\right)^{n-2} \Upsilon(1\cdots n),
\label{eq:canonical-Upsilon} \\
\Xi^s(1\cdots n) &\rightarrow
    \left(-\frac{ig}{\sqrt 2}\right)^{n-2} \Xi^s(1\cdots n).
\label{eq:canonical-Xi}
\end{align}
With the off-shell definitions \eqref{eq:lc-4d-spinorbrackets},
the vertices are then precisely the Parke-Taylor MHV amplitudes:
\begin{equation}
\label{eq:pta}
V^s(1\cdots n) = g^{n-2}{\langle 1\:s\rangle^4\over\langle
1\:2\rangle\langle 2\:3\rangle\cdots\langle n-1,n\rangle\langle
n\:1\rangle}.
\end{equation}

\section{Tree-Level Equivalence Theorem Evasion}
\label{sec:treelevel-etviol}

Now we have enough tools to obtain the tree-level $(\mpp)$ amplitude, which
cannot be constructed from the vertices of the theory. We explain that this
arises from a evasion of the equivalence theorem.

\subsection{On LSZ reduction and scattering amplitudes}
\label{ssec:onlsz}

Scattering amplitudes are formed by the application of
LSZ reduction to correlation functions of the $A$ fields. For outgoing momenta
$\{p_i\}$ and helicities $\{h_i\}$,
\begin{equation}
\label{eq:lsz}
\langle p_1^{h_1},\dots,p_n^{h_n} \rvert S \lvert 0 \rangle
    = (-i)^n \:
      \lim_{p_i^2 \rightarrow 0}
      p_1^2 \cdots p_n^2 \: \langle
          {E^{\mu_1}_{h_1}} A_{\mu_1} \cdots
      {E^{\mu_n}_{h_n}} A_{\mu_n} \rangle.
\end{equation}
The $E_{h_i}^{\mu_i}$ are polarisation vectors in four dimensions
using the outgoing helicity convention, hence no complex
conjugation. The polarisations are given (in the spinor helicity formalism)
by
\begin{equation}
E_+ = \sqrt 2 \frac{\eta\,\tilde\lambda}{\langle\eta\:\lambda\rangle}
\quad\text{and}\quad
E_- = \sqrt 2 \frac{\lambda\,\tilde\eta}{[\eta\:\lambda]}
\end{equation}
where (up to an unimportant phase) $\eta$ and $\tilde\eta$ are
defined as below \eqref{eq:lc-4d-spinorbrackets}. Then $E_+=\bar
E_-=-1$, and by considering the invariant
\eqref{eq:lc-4d-invariant}, we see that each $+$ ($-$) external
state gluon on the left-hand side of \eqref{eq:lsz} is associated
with an $\bar A$ ($A$) field in the correlator on the right-hand
side. Note that the $+$ ($-$) external state is associated with
the expected  $A$ ($\bar A$) field only after `amputating' the
corresponding propagator on the right-hand side of \eqref{eq:lsz}.

\begin{figure}[t]
  \begin{align*}
    \begin{matrix}\begin{picture}(60,52)
        \SetOffset(30,24)
        \Photon(-30,0)(-2,0){2}{4}
        \Line(0,0)(30,0)
        \BCirc(0,0){2}
        \Text(-30,3)[bl]{$+$}
        \Text(30,3)[br]{$+$}
      \end{picture}\end{matrix}&=1
    &
    \begin{matrix}\begin{picture}(85,52)
        \SetOffset(30,24)
        \Photon(-30,0)(-2,0){2}{4}
        \Line(0,0)(30,15)
        \Line(0,0)(30,-15)
        \BCirc(0,0){2}
        \LongArrow(-5,-7)(-30,-7)
        \LongArrow(8.2111,8.5777)(28.2111,18.5777)
        \LongArrow(8.2111,-8.5777)(28.2111,-18.5777)
        \Text(-30,3)[bl]{$+$}
        \Text(32,15)[cl]{$+$}
        \Text(32,-15)[cl]{$+$}
        \Text(-17.5,-10)[tc]{$p_1$}
        \Text(22,16)[br]{$p_2$}
        \Text(22,-16)[tr]{$p_3$}
      \end{picture}\end{matrix}&=\Upsilon(123)
    \\
    \begin{matrix}\begin{picture}(60,52)
        \SetOffset(30,24)
        \Photon(-30,0)(-2,0){2}{4}
        \Line(0,0)(30,0)
        \BCirc(0,0){2}
        \Text(-30,3)[bl]{$-$}
        \Text(30,3)[br]{$-$}
      \end{picture}\end{matrix}&=1
    &
    \begin{matrix}\begin{picture}(85,52)
        \SetOffset(30,24)
        \Photon(-30,0)(-2,0){2}{4}
        \Line(0,0)(30,15)
        \Line(0,0)(30,-15)
        \BCirc(0,0){2}
        \Text(-30,3)[bl]{$-$}
        \Text(32,15)[cl]{$-$}
        \Text(32,-15)[cl]{$+$}
      \end{picture}\end{matrix}&=-\dfrac{\hat 2}{\hat 1}\Xi^1(123)
    \\
    & &
    \begin{matrix}\begin{picture}(85,52)
        \SetOffset(30,24)
        \Photon(-30,0)(-2,0){2}{4}
        \Line(0,0)(30,15)
        \Line(0,0)(30,-15)
        \BCirc(0,0){2}
        \Text(-30,3)[bl]{$-$}
        \Text(32,15)[cl]{$+$}
        \Text(32,-15)[cl]{$-$}
      \end{picture}\end{matrix}&=-\dfrac{\hat 3}{\hat 1}\Xi^2(123)
  \end{align*}
  \caption{The MHV completion vertices:
    graphical representations of the $\Upsilon$ and $\Xi$ coefficients
    of the series expansion of $\gA$ and $\bar \gA$ (shown up to
    $\mathcal{O}({\gB^2})$). The wavy lines with a $+$($-$) denote insertions
    of $\gA$($\bar\gA$) operators in correlation functions; $\gB$ and
    $\bar\gB$ attach to the straight lines.  }
  \label{fig:feyn-UpsilonXi-4d}
\end{figure}

A generic correlation function of the $A$ fields may be written schematically
in momentum space as 
\[
\langle \cdots A(p) \cdots \bar A(q) \cdots \rangle.
\]
However,  it is now the $B$ fields which propagate. Therefore, we
must regard $A$ and $\bar A$ above as functions of $B$ and $\bar
B$ and make the replacements from the series; again, schematically
we can write this as (neglecting the field normalisation factors from
\eqref{eq:canonical-Upsilon} and \eqref{eq:canonical-Xi})
\begin{equation}
\label{eq:Wick}
\langle
    \cdots
    \biggl(
        \sum_n \Upsilon_{p2\cdots n}B_{\bar 2}\cdots B_{\bar n}
    \biggr)\cdots\biggl( -
        \sum_{n,s} \frac{\hat s}{\hat q} \Xi^{s-1}_{q2\cdots n}
            B_{\bar 2}\cdots\bar B_{\bar s}\cdots B_{\bar n}
    \biggr)\cdots
\rangle.
\end{equation}
Order-by-order, we take Wick contractions between the $B$ field operators
with its propagator. This naturally lends itself to a Feynman graph
representation where we have vertices for the $\Upsilon$ and $\Xi$
coefficients, the first few of which are shown in
fig.~\ref{fig:feyn-UpsilonXi-4d}. We refer to these vertices as ``MHV
completion vertices'' and graphs built from them as ``MHV completion
graphs'' since they allow the construction of amplitudes otherwise absent
from the theory. (Note that the figure shows the vertices appropriate for
$\gA$, $\gB$, etc. so that the normalisation factors of
\eqref{eq:canonical-Upsilon} and \eqref{eq:canonical-Xi} above can be
omitted for clarity.)

\subsection{Tree-level $(\mpp)$ amplitude}
\label{ssec:treelevel-mpp}

\begin{figure}[t]
  \centering \subfigure{
    \begin{picture}(100,80) \SetOffset(50,27) \Photon(0,27)(0,40){2}{2}
      \Photon(23.3827,-13.5)(34.6410,-20){2}{2}
      \Photon(-23.3827,-13.5)(-34.6410,-20){2}{2} \Line(0,25)(21.6506,-12.5)
      \Line(0,25)(-21.6506,-12.5) \BCirc(0,25){2} \BCirc(21.6506,-12.5){2}
      \BCirc(-21.6506,-12.5){2} \Text(0,43)[bc]{$1^+$}
      \Text(35.6410,-19)[tl]{$2^-$} \Text(-35.6410,-19)[tr]{$3^-$}
      \Text(-4,24)[tr]{$+$} \Text(4,24)[tl]{$+$} \Text(-19.0,-8.5)[br]{$-$}
      \Text(19.0,-8.5)[bl]{$-$}
    \end{picture}
    \label{fig:feyn-ppm-1}
  }\quad \subfigure{
    \begin{picture}(100,80) \SetOffset(50,27) \Photon(0,27)(0,40){2}{2}
      \Photon(23.3827,-13.5)(34.6410,-20){2}{2}
      \Photon(-23.3827,-13.5)(-34.6410,-20){2}{2} \Line(0,25)(21.6506,-12.5)
      \Line(21.6506,-12.5)(-21.6506,-12.5) \BCirc(0,25){2}
      \BCirc(21.6506,-12.5){2} \BCirc(-21.6506,-12.5){2}
      \Text(4,24)[tl]{$+$} \Text(19.0,-8.5)[bl]{$-$}
      \Text(-19,-13.5)[tl]{$-$} \Text(19,-13.5)[tr]{$+$}
    \end{picture}
    \label{fig:feyn-ppm-2}
  }\quad \subfigure{
    \begin{picture}(100,100) \SetOffset(50,27) \Photon(0,27)(0,40){2}{2}
      \Photon(23.3827,-13.5)(34.6410,-20){2}{2}
      \Photon(-23.3827,-13.5)(-34.6410,-20){2}{2}
      \Line(0,25)(-21.6506,-12.5) \Line(21.6506,-12.5)(-21.6506,-12.5)
      \BCirc(0,25){2} \BCirc(21.6506,-12.5){2} \BCirc(-21.6506,-12.5){2}
      \Text(-4,24)[tr]{$+$} \Text(-19.0,-8.5)[br]{$-$}
      \Text(-19,-13.5)[tl]{$+$} \Text(19,-13.5)[tr]{$-$}
    \end{picture}
    \label{fig:feyn-ppm-3}
  }
  \caption{Contributions to the tree-level $(\mpp)$ amplitude, before
    applying LSZ reduction.}
  \label{fig:feyn-ppm}
\end{figure}
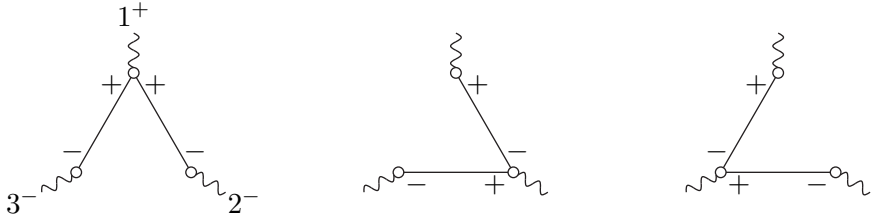

We obtain the amplitude by amputating the $\langle A \bar A \bar A
\rangle$ correlation function, whose tree-level contributions are
shown in fig.~\ref{fig:feyn-ppm}. Hence, we need to take the limit
as $p_1^2, p_2^2, p_3^2 \rightarrow 0$ of
\begin{equation}
\begin{split}
A(1^-,2^+,3^+) &=
    -i p_1^2 p_2^2 p_3^2 \left\{
        \frac1{p_2^2} \frac1{p_3^2} \Upsilon(123)
        - \frac1{p_3^2} \frac1{p_1^2} \frac{\hat 1}{\hat 2} \Xi^2(231)
        - \frac1{p_1^2} \frac1{p_2^2} \frac{\hat 1}{\hat 3} \Xi^1(312)
        \right\} \left(-\frac{ig}{\sqrt 2}\right) \\
    &=
       \frac{ig}{\sqrt 2} \frac{\hat 1^2}{(2\:3)} \left(
           \frac{p_1^2}{\hat 1} + \frac{p_2^2}{\hat 2} + \frac{p_3^2}{\hat 3}
       \right)
\end{split}
\label{eq:feyn-ppm-raw}
\end{equation}
In the first line, the leading factor of $-i$ comes from an un-cancelled inverse
propagator, and the final factor from the application of
\eqref{eq:canonical-Upsilon}.

The third factor on the second line can further be simplified by applying
\eqref{eq:sum-bilinears}, reducing further to
\begin{equation}
\begin{split}
A(1^-,2^+,3^+)
&= i g \sqrt 2 \frac{\hat 1}{\hat 2 \hat 3} \{ 2\:3 \} \\
&= i g \frac{\phantom{{}^3}[2\:3]^3}{[3\:1][1\:2]},
\end{split}
\label{eq:feyn-ppm-raw-end}
\end{equation}
the expected \mhvbar\ amplitude.

As is well known, in order for this to be non-vanishing in the
on-shell limit, we need to work with complex momenta or $(2,2)$
signature. However, it is noteworthy that off mass shell it actually
coincides with the CSW prescription.

\subsection{The origin of equivalence theorem evasion}
\label{ssec:etviol-origin}

It may be useful to illustrate the general mechanism behind the S-matrix 
equivalence
theorem \cite{Itzykson:1980rh} with a toy scalar model. We compute
correlation functions from the model's partition function by adding a source
term $\int d^Dx\:j\phi$. Now write the action in terms of a new field
variable $\phi'$ given implicitly by the invertible transformation $\phi =
f(\phi',\partial_\mu \phi',\partial_\mu\partial_\nu\phi',\dots)$ where $f$
is a \emph{regular} function of $\phi'$ and its derivatives.  (If it has a
non-unit jacobian we can ignore it for the purposes of this discussion.)
Upon taking derivatives with respect to $j$, we see that additional terms of
$\phi'^2$ and higher powers are also pulled down from the exponential. An
insertion of $\phi'^n(x)$ will connect to $n$ propagators, whose momenta
sum to that associated with $x$ by the Fourier transform. Unlike when
$n=1$, these propagators do not, in general, cancel the inverse propagator
from LSZ reduction and thus vanish in the on-shell limit. Hence, we can
truncate the source term to $\int d^Dx\:j\phi'$. At the quantum level,
self-energy-like terms can be made from insertions of $\phi'^n(x)$, but this
will alter scattering amplitudes by at most a wavefunction renormalisation
(as has been discussed in this context in ref. \cite{Ettle:2006bw}).

The canonical transformation considered here is not local  within the
quantisation surface but it is local in  light-cone time. This vestige of
locality is usually  sufficient to make the theorem applicable at tree-level
because  the canonical transformation contains no terms in  $\check{p}$ and
so cannot produce a factor of $1/p^2$ which  contains $\check{p}$ in the
denominator. It is therefore  remarkable that in certain cases the theorem
is circumvented. This  can happen when terms collect so that the factor
$\sum_j  p^2_j/\hat j$ is formed, as in \eqref{eq:feyn-ppm-raw}. By 
\eqref{eq:sum-omega} such terms are independent of  $\check{p}$ and can be
cancelled by the restricted  non-locality introduced through
\eqref{eq:Upsilon-4d-rr}. As we will see, the same features are responsible
for the  recovery of the all-$+$ amplitudes at one loop.

\section{The $D$-Dimensional Canonical MHV Lagrangian}
\label{sec:dimreg-mhvl}

The treatment of quantum corrections to amplitudes in the canonical MHV
lagrangian formalism will require that the theory be regulated, and we will
do so by dimensional regularisation. It turns out that we can then apply the
canonical transformation procedure essentially as before, save for the fact
that pieces outside four dimensions result in much richer structure and
hence more complicated MHV rules.

\subsection{Light-cone Yang-Mills in $D$ dimensions}
\label{ssec:lc-Dd}

We write the co-ordinates in $D=4-2\epsilon$ dimensions as:
\begin{align*}
       x^0 &= \tfrac1{\sqrt 2}(t-x^{D-1}), &
       z^I &= \tfrac1{\sqrt 2}(x^{2I-1}+ix^{2I}),\\
x^{\bar 0} &= \tfrac1{\sqrt 2}(t+x^{D-1}), &
  \bar z^I &= \tfrac1{\sqrt 2}(x^{2I-1}-ix^{2I}),
\end{align*}
where the index $I$ runs over the $\tfrac12(2-2\epsilon)$ pairs of
transverse directions.
In these
co-ordinates, the metric takes block diagonal form with non-zero
components $g_{0\bar0}=g_{\bar0 0}=1$, $g_{z^I\bar z^J}=g_{\bar
z^I z^J}=-\delta_{IJ}$. Again, we introduce a more compact
notation for the components of 1-forms and momenta, for which we
write $(p_0, p_{\bar 0}, p_{z^I}, p_{\bar z^J}) \equiv (\check{p},
\hat p, p_I, \bar p_I)$, with $(\check{n}, \hat n, n_I, \bar n_I)$
for momenta labelled by a number.

The reason we make this choice of basis is that it will lead us
again to a lagrangian with the structure \eqref{eq:cmhvl-4d} and
thus inherit some of the simplicity of MHV rules in four space-time
dimensions, for example the tree-level properties that the first
non-vanishing amplitudes are MHV amplitudes, these coinciding with
the lagrangian vertices, that NMHV amplitudes are constructed by
joining precisely two such vertices together by the propagator and
so on.

In these co-ordinates, the invariant becomes
\begin{equation}
A \cdot B = \check{A}\,\hat B + \hat A\,\check{B} - A_I \bar B_I
    - \bar A_I B_I,
\label{eq:paralc-invariant}
\end{equation}
where we have assumed the summation convention that a repeated capital Roman
index in a product is summed over $1,\dots,1-\epsilon$. The bilinears of
\eqref{eq:lc-4d-bilinears} become
\begin{equation}
\label{eq:lc-d-bilinears}
  (1\:2)_I := \hat 1 2_I - \hat 2 1_I, \quad
\{1\:2\}_I := \hat 1 \bar 2_I - \hat 2 \bar 1_I.
\end{equation}
They amount to our $D$ dimensional generalisation of the familiar
spinor brackets \eqref{eq:lc-4d-spinorbrackets}.
Scalar products between these will often be
shortened to
\[
(1\:2)\mdot\{1\:2\} \equiv (1\:2)_I \{1\:2\}_I,
\]
where the dot is obviously redundant when the bilinears are purely
four-dimensional.

The Yang-Mills action is written as before in Minkowski co-ordinates as
\begin{equation}
S=\frac 1{2g^2} \int d^Dx\:\tr \gF^{\mu\nu}\gF_{\mu\nu}.
\label{eq:ym-action-Dd}
\end{equation}
The field-strength tensor and group generators are defined as
before in \eqref{eq:ym-defs} and \eqref{eq:ym-generators}. The
quantisation procedure is similar to that in four dimensions. It
takes place on surfaces $\Sigma$ with normal $\mu =
(1,0,\dots,0,1) / \sqrt 2$ (\ie\ of constant $x^0$) in Minkowski
co-ordinates, subject to the axial gauge condition $\mu~\cdot~\gA
= \hat\gA = 0$. We integrate $\check\gA$ out of the lagrangian and
are left with a $D$-dimensional light-cone action in the form
\eqref{eq:ym-lc4d-action}, where now however
\newcommand{\trintx}{\tr\int_\Sigma d^{D-1}\vec{x}\:}
\begin{align}
  L^{-+} &= \phantom{-} \trintx
    \gA_I (\check\partial\hat\partial - \partial_J\bar\partial_J) \bar \gA_I,
\label{eq:hyperlc-mp}
\\
 L^{-++} &=          -  \trintx
    (  \bar\partial_I \gA_J [\hat\partial^{-1} \gA_I, \hat\partial \bar \gA_J]
     + \bar\partial_I \bar \gA_J [\hat\partial^{-1} \gA_I, \hat\partial \gA_J]
     ),
\label{eq:hyperlc-mpp}
\\
 L^{--+} &=          -  \trintx
    (  \partial_I \gA_J [\hat\partial^{-1} \bar \gA_I, \hat\partial \bar \gA_J]
     + \partial_I \bar \gA_J [\hat\partial^{-1} \bar \gA_I, \hat\partial \gA_J]
     ),
\label{eq:hyperlc-mmp}
\\
\begin{split}
L^{'--++} &=          -  \trintx \biggl(
      \hphantom{+\,}
        \frac 14 [\hat\partial \gA_I,\bar \gA_I] \:\hat\partial^{-2}\:
        [\hat\partial \gA_J,\bar \gA_J]  \\
    & \hphantom{=-\trintx\biggl(} +
        \frac 12 [\hat\partial \gA_I,\bar \gA_I] \:\hat\partial^{-2}\:
        [\hat\partial \bar \gA_J, \gA_J] \\
    & \hphantom{=-\trintx\biggl(} -
        \frac 14 [\hat\partial \bar \gA_I, \gA_I] \:\hat\partial^{-2}\:
        [\hat\partial \bar \gA_J, \gA_J] \\
    & \hphantom{=-\trintx\biggl(} -
        \frac 14 [ \gA_I, \gA_J ] [ \bar \gA_I, \bar \gA_J ]
    - \frac 14 [ \gA_I, \bar \gA_J ] [ \bar \gA_I, \gA_J ] \biggr).
\end{split}
\label{eq:hyperlc-mmpp}
\end{align}
It may be shown with integration by parts that these expressions reduce in four
dimensions to those of Section \ref{ssec:lightcone-ym-4d}.

\subsection{The transformation}
\label{ssec:mhvl-trans}

We will now specify the change of field variables from $\gA$ and $\bar\gA$
to $\gB$ and $\bar\gB$.  From \eqref{eq:hyperlc-mp}, we see that the
momentum conjugate to $\gA_I$ is $\Pi_I(x)=-\hat\partial\bar\gA_I(x)$, and
as such
\begin{equation}
  \pD\gA\:\pD\Pi \equiv \prod_{x,I} d\gA_I(x)\:d\Pi_I(x)
  \label{eq:pathint-measure}
\end{equation}
is proportional (up to a constant) to the path integral measure
$\pD\gA\:\pD\bar\gA$; therefore a canonical field transformation will have
unit jacobian.  We can implement this by requiring that $\gA$ be a
functional of $\gB$ alone, subject to either of the following equivalent
conditions:
\begin{gather}
  \hat\partial \bar\gA^a_I(x^0,\vec x) = \int_\Sigma d^{D-1}\vec y \:
  \frac{\delta\gB^b_J(x^0,\vec y)}{\delta\gA^a_I(x^0,\vec x)} \hat\partial
  \bar\gB^b_J(x^0,\vec y),
  \label{eq:momentum-transform} \\
  \tr \int_\Sigma d^{D-1}\vec x\:\check\partial\gA_I\,\hat\partial\bar\gA_I
  = \tr \int_\Sigma d^{D-1}\vec
  x\:\check\partial\gB_I\,\hat\partial\bar\gB_I.
  \label{eq:second-canon-invar}
\end{gather}

Again, working in momentum space on the quantisation surface, we express
$\gA$ as a series in $\gB$, but this time the series coefficients carry
extra indices for the transverse directions:
\begin{equation}
  \label{eq:hypermft-A-series}
  \gA_{I_1}(\vec p_1) = \sum_{n=2}^{\infty} \int_{2\cdots n}
  \Upsilon_{I_1\cdots I_n}(1\cdots n) \:
  \gB_{I_2}(-\vec p_2) \cdots \gB_{I_n}(-\vec p_n)
\end{equation}
where $\Upsilon_{IJ}(12) = \delta(\vec p_1+\vec p_2) \delta_{IJ}$.  The
integral short-hand here is defined by
\[
\int_{1\cdots n} = \prod_{k=1}^n \frac1{(2\pi)^{3-2\epsilon}} \int d\hat k
\prod_{I=1}^{1-\epsilon} dk_I d\bar k_I
\]
and for later use we introduce the $\delta$-function stripped form of a
coefficient, given (as the first factor on the right-hand side) by
\begin{equation}
  \Upsilon_{I_1\cdots I_n}(1\cdots n) =
  \Upsilon(1^{I_1}\cdots n^{I_n}) \:
  (2\pi)^{3-2\epsilon} \delta(\vec p_1 + \cdots + \vec p_n)
\end{equation}
and similarly for the other vertices $\Xi$, $V$ and $W$, defined below. They
should only be considered to be defined when the sum of their momentum
arguments is $0$. Repeated transverse indices in the superscripts are also
subject to the summation convention. For convenience, we will often also
subsume the index into the momentum label when the association is obvious
(e.g.\ $\Upsilon(1^{I_1}\cdots n^{I_n}) \rightarrow \Upsilon(1\cdots n)$
above).

The canonical transformation removes the $(\mpp)$ terms from the lagrangian
by absorbing them into the kinetic term for $\gB$:
\begin{equation}
  L^{-+}[\gA] + L^{-++}[\gA] = L^{-+}[\gB].
  \label{eq:AB-implicit}
\end{equation}
Briefly delving into momentum space on the quantisation surface, it is seen
that the term on the right-hand side of \eqref{eq:AB-implicit} furnishes the
tree-level propagator
\begin{equation}
  \langle \gB_I \bar\gB_J \rangle = -\frac{ig^2}{2p^2}\delta_{IJ}.
\end{equation}
Similarly, from the quantisation surface Fourier transform of
\eqref{eq:hyperlc-mpp}, expanding the commutator and re-labelling leads us
to
\begin{equation}
  L^{-++} = \tr\int_{123}
  \bar V^2_{IJK}(123) \: \gA_I(\bar1) \gA_J(\bar2) \bar\gA_K(\bar3)
  \label{eq:hyperlc-mom-mpp}
\end{equation}
where
\begin{equation}
  \bar V^2(1^I2^J3^K) = i \left(
    \frac{\{3\:1\}_J \delta_{KI}}{\hat 2}
    + \frac{\{2\:3\}_I \delta_{JK}}{\hat 1}
  \right).
\label{eq:Vbar2}
\end{equation}
It obviously follows from \eqref{eq:AB-implicit} and the
light-cone lagrangian that this is
the $D$ dimensional equivalent of
the $(\ppm)$ \mhvbar\ vertex for the $\gA$ field, and this is
reflected in our choice of notation.

The remaining pieces of the lagrangian,
\eqref{eq:hyperlc-mmp} and \eqref{eq:hyperlc-mmpp}, form MHV vertices in
$4-2\epsilon$ dimensions, as explained in the next section.  To obtain the
$\Upsilon$ coefficients, we take the explicit expression of
\eqref{eq:AB-implicit} and use \eqref{eq:momentum-transform} and
\eqref{eq:second-canon-invar} to further reduce it to
\begin{equation}
  \left\{ \frac{\partial \mdot \bar\partial}{\hat \partial} \gA_I
    - [\bar\partial_J \gA_I, \hat\partial^{-1} \gA_J]
    - \frac{\bar\partial_J}{\hat\partial} [\hat\partial^{-1} \gA_J,
    \hat\partial \gA_I] \right\} (\vec x)
  = \int_\Sigma d^3\vec y \:
  \frac{\delta\gA_I(\vec x)}{\delta\gB^b_J(\vec y)}
  \left(\frac{\partial \mdot \bar\partial}
    {\hat \partial}\right)_{\vec y}
  \gB^b_J(\vec y).
  \label{eq:AB-explicit}
\end{equation}
By again transforming to momentum space and substituting the series
expansion for $\gA$ into both sides of \eqref{eq:AB-explicit} above,
carefully rearranging the fields, and comparing terms order-by-order in
$\gB$, we extract successive $\Upsilon$ coefficients. At
$\mathcal{O}(\gB^2)$, one finds
\begin{align}
  \label{eq:upsilon-3}
  \Upsilon(1^I2^J3^K) &=
  \frac{i}{(2\:3)\mdot\{2\:3\}}
  (\hat 2 \{2\:3\}_K \delta_{IJ} + \hat 3 \{2\:3\}_J \delta_{KI}) \\
  &= -\frac{1}{\hat{1}} \frac{\bar{V}^2(2^J 3^K 1^I)}
                              {(\Omega_1 + \Omega_2 + \Omega_3)} 
			      \notag \\
  &= \frac{2}{\hat 1} \frac{\bar{V}^2(2^J 3^K 1^I)}
      {p_1^2/\hat 1 + p_2^2/\hat 2 + p_3^2/\hat 3}
\end{align}
and for compactness we have defined
\[
\Omega_p := \frac{p_I \bar p_I}{\hat p}.
\]
By continuation of this process, we arrive at the following recursion
relation for $\Upsilon$:
\begin{equation}
  \begin{split}
    \Upsilon(1\cdots n) = - \frac{i}{\sum_{i=1}^n \Omega_i} \sum_{j=2}^{n-1}
    & [ \zeta(1,P_{j+1,n}^B,P_{2j}^A)
    - \zeta(1,P_{2j}^A,P_{j+1,n}^B) ] \\
    & \times \Upsilon(-^A,2,\dots,j) \Upsilon(-^B,j+1,\dots,n),
  \end{split}
  \label{eq:upsilon-rr}
\end{equation}
where $P_{ij}:=p_i + p_{i+1} \cdots + p_j$ and we have introduced the symbol
\[
\zeta(1^I2^J3^K) := \frac{\{2\:3\}_K \delta_{IJ}}{\hat 3(\hat 2 + \hat 3)}.
\]
An alternative form for this expression follows from \eqref{eq:AB-implicit}:
\begin{equation}
  \Upsilon(1\cdots n) = -
  \frac{1}{\hat1 \sum_{i=1}^n \Omega_i} \sum_{j=2}^{n-1}
  \bar V^2(P^A_{2j},P^B_{j+1,n},1)
  \Upsilon(-^A,2,\dots,j)
  \Upsilon(-^B,j+1,\dots,n),
  \label{eq:upsilon-rr-x}
\end{equation}

Of course it is no accident that the recurrence relation and its
solutions can be expressed in terms of $\bar V^2$ in this way. The
canonical transformation that absorbs this interaction into the
kinetic term in \eqref{eq:AB-implicit} can be performed for any
choice of $\bar V^2$. This will be important later on, since it
clarifies how the amplitudes involving the missing $\bar V^2$
vertex are recovered, and also demonstrates that the mechanism of
recovery is generic in any theory where a canonical transformation
is used to absorb a three point vertex into the kinetic term as in
\eqref{eq:AB-implicit}.

A useful particular case of this formula is
\begin{equation}
\begin{split}
  \Upsilon(1234) = \frac{1}{\hat 1 \sum_{i=1}^4 \Omega_i}
  \Biggl\{& \bar V^2(2, \bar 5^A, 1) \frac{1}{\hat 5(\Omega_5 +
    \Omega_3 + \Omega_4)} \bar V^2(3, 4, 5^A) \\
  + & \bar V^2(\bar 5^A, 4, 1) \frac{1}{\hat 5 (\Omega_5 +
    \Omega_2 + \Omega_3)} \bar V^2(2, 3, 5^A)\Biggr\}.
\end{split}
\label{eq:upsilon1234}
\end{equation}
Note that here (and throughout) $5$ is a dummy momentum with scope limited
to each term, and that its value should be taken to be the negative of the
sum of the other arguments that accompany it a vertex.

Differentiating \eqref{eq:hypermft-A-series} with respect to $\gB$ and
inserting the inverse into \eqref{eq:momentum-transform} suggests a series
expansion for $\bar \gA$ of the form
\begin{equation}
  \label{eq:hypermft-Abar-series}
  \bar\gA_{I_1}(-\vec p_1) = \sum_{n=2}^{\infty} \sum_{s=2}^n \int_{2\cdots n}
  \frac{\hat s}{\hat 1} \Xi^{s-1}_{I_1\cdots I_n}(\bar12\cdots n) \:
  \gB_{I_2}(-\vec p_2)
  \cdots \bar\gB_{I_s}(-\vec p_s) \cdots
  \gB_{I_n}(-\vec p_n).
\end{equation}
Now, inserting \eqref{eq:hypermft-A-series} and
\eqref{eq:hypermft-Abar-series} into \eqref{eq:second-canon-invar}, and then
comparing coefficients order-by-order in $\gB$, we may obtain expressions
for the $\Xi$ coefficients in terms of other $\Xi$s and $\Upsilon$s of lower
order. The results
\begin{equation}
  \label{eq:xi-3pt}
  \Xi^1(1^I2^J3^K) = - \Upsilon(2^J3^K1^I)
  \quad\text{and}\quad
  \Xi^2(1^I2^J3^K) = - \Upsilon(3^K1^I2^J)
\end{equation}
will be of particular relevance to the forthcoming. By careful examination
of the expansion of \eqref{eq:second-canon-invar} at the $(n-1)^{\rm th}$
order in $\gB$, one finds that this recursion relation
\begin{equation}
  \begin{split}
    \Xi^s(1\cdots n) = - \sum_{r=\max(2,4-s)}^{n+1-s}
    \sum_{m=\max(r,3)}^{r+s-1}
    &   \Upsilon(-^A,3-r,\dots,m+1-r) \\
    & \times \Xi^{r+s-m}(-^A,m+2-r,\dots,2-r).
  \end{split}
  \label{eq:xi-rr}
\end{equation}
holds, given that $\Xi(1^I2^J)=\delta_{IJ}$.

As in section \ref{ssec:onlsz}, we introduce a diagrammatic
representation of $\Upsilon$ and $\Xi$ in the form of $D$-dimensional MHV
completion vertices, the first few of which are shown in
fig.~\ref{fig:feyn-UpsilonXi}. We note that the process of deriving
$\Upsilon$ and $\Xi$ in $4-2\epsilon$ dimensions differs only from that in
four dimensions by the presence of extra transverse indices, which are seen
to each ride alongside (and can therefore be built into) a momentum index.
It is therefore not surprising that the relationship between $\Upsilon$ and
$\Xi$ is, from this point of view, identical to that in four dimensions.
\begin{figure}[h]
  \begin{align*}
    \begin{matrix}\begin{picture}(60,52)
        \SetOffset(30,24)
        \Photon(-30,0)(-2,0){2}{4}
        \Line(0,0)(30,0)
        \BCirc(0,0){2}
        \Text(-30,3)[bl]{$+,I$}
        \Text(30,3)[br]{$+,J$}
      \end{picture}\end{matrix}&=\delta_{IJ}
    &
    \begin{matrix}\begin{picture}(85,52)
        \SetOffset(30,24)
        \Photon(-30,0)(-2,0){2}{4}
        \Line(0,0)(30,15)
        \Line(0,0)(30,-15)
        \BCirc(0,0){2}
        \LongArrow(-5,-7)(-30,-7)
        \LongArrow(8.2111,8.5777)(28.2111,18.5777)
        \LongArrow(8.2111,-8.5777)(28.2111,-18.5777)
        \Text(-30,3)[bl]{$+,I$}
        \Text(32,15)[cl]{$+,J$}
        \Text(32,-15)[cl]{$+,K$}
        \Text(-17.5,-10)[tc]{$p_1$}
        \Text(22,16)[br]{$p_2$}
        \Text(22,-16)[tr]{$p_3$}
      \end{picture}\end{matrix}&=\Upsilon(1^I2^J3^K)
    \\
    \begin{matrix}\begin{picture}(60,52)
        \SetOffset(30,24)
        \Photon(-30,0)(-2,0){2}{4}
        \Line(0,0)(30,0)
        \BCirc(0,0){2}
        \Text(-30,3)[bl]{$-,I$}
        \Text(30,3)[br]{$-,J$}
      \end{picture}\end{matrix}&=\delta_{IJ}
    &
    \begin{matrix}\begin{picture}(85,52)
        \SetOffset(30,24)
        \Photon(-30,0)(-2,0){2}{4}
        \Line(0,0)(30,15)
        \Line(0,0)(30,-15)
        \BCirc(0,0){2}
        \Text(-30,3)[bl]{$-,I$}
        \Text(32,15)[cl]{$-,J$}
        \Text(32,-15)[cl]{$+,K$}
      \end{picture}\end{matrix}&=-\dfrac{\hat 2}{\hat 1}\Xi^1(1^I2^J3^K)
    \\
    & &
    \begin{matrix}\begin{picture}(85,52)
        \SetOffset(30,24)
        \Photon(-30,0)(-2,0){2}{4}
        \Line(0,0)(30,15)
        \Line(0,0)(30,-15)
        \BCirc(0,0){2}
        \Text(-30,3)[bl]{$-,I$}
        \Text(32,15)[cl]{$+,J$}
        \Text(32,-15)[cl]{$-,K$}
      \end{picture}\end{matrix}&=-\dfrac{\hat 3}{\hat 1}\Xi^2(1^I2^J3^K)
  \end{align*}
  \caption{ $D$-dimensional $\Upsilon$ and $\Xi$ MHV
    completion vertices. }
  \label{fig:feyn-UpsilonXi}
\end{figure}

\subsection{The hyper-MHV rules}
\label{ssec:hypermhv}

We will now extract the $4-2\epsilon$-dimensional generalisations of the
three and four gluon MHV amplitudes. The interaction part of the lagrangian
takes the same form as \eqref{eq:V} except that the vertices carry
polarisation indices to contract into the corresponding $\gB$s and
$\bar\gB$s.  The Feynman rule for a particular $\gB$ vertex is thus
$4iV^s(1^{I_1}\cdots n^{I_n})/g^2$, and this follows from the definition as the
sum of all contractions of external lines into the term in the action with
the matching colour factor, while accounting for the cyclic symmetry of the
trace.

The three-point MHV vertex can be obtain in a manner almost identical to how
\eqref{eq:Vbar2} was derived. In quantisation surface momentum-space,
\eqref{eq:hyperlc-mmp} reads
\begin{equation}
L^{--+} = \tr\int_{123}
    V^2_{IJK}(123) \: \bar\gA_I(\bar1) \bar\gA_J(\bar2) \gA_K(\bar3)
\label{eq:hyperlc-mom-mmp}
\end{equation}
where
\begin{equation}
V^2(1^I2^J3^K) = i \left(
    \frac{(3\:1)_J \delta_{KI}}{\hat 2}
  + \frac{(2\:3)_I \delta_{JK}}{\hat 1}
\right).
\end{equation}
Since $\gA=\gB$ and $\bar\gA=\bar\gB$ to leading order, upon
substituting \eqref{eq:hypermft-A-series} and
\eqref{eq:hypermft-Abar-series} into \eqref{eq:hyperlc-mom-mmp},
we immediately see that $V^2(1^I2^J3^K)$ is the
$\bar\gB_I(1)\bar\gB_J(2)\gB_K(3)$ colour-ordered vertex.

We note that $\bar\gB\bar\gB\gB\gB$ and $\bar\gB\gB\bar\gB\gB$
colour-ordered vertices receive contributions from $L^{--+}[\gA]$ and
$L^{'--++}[\gA]$. Upon writing the latter in momentum-space, we have
\begin{equation}
  \begin{split}
    L^{'--++} = \tr\int_{1234} \bigl\{ & W^2_{IJKL}(1234) \:
    \bar\gA_I(\bar1) \bar\gA_J(\bar2) \gA_K(\bar3) \gA_L(\bar4) \\
    & + W^3_{IJKL}(1234) \: \bar\gA_I(\bar1) \gA_J(\bar2) \bar\gA_K(\bar3)
    \gA_L(\bar4) \bigr\}
  \end{split}
  \label{eq:hyperlc-mom-mmpp}
\end{equation}
where
\begin{align}
  W^2(1^I2^J3^K4^L) &= \delta_{IK}\delta_{JL} + \delta_{IL}\delta_{JK}
  \frac{\hat 1 \hat 2 + \hat 3 \hat 4}{(\hat 1 + \hat 4)^2}, \\
  W^3(1^I2^J3^K4^L) &= \frac12 \left( \delta_{IL}\delta_{JK} \frac{\hat 1
      \hat 2 + \hat 3 \hat 4}{(\hat 1 + \hat 4)^2} + \delta_{IJ}\delta_{KL}
    \frac{\hat 1 \hat 4 + \hat 2 \hat 3}{(\hat 1 + \hat 2)^2} \right).
\end{align}
We substitute \eqref{eq:hypermft-A-series} and
\eqref{eq:hypermft-Abar-series} into $L^{--+}[\gA] + L^{'--++}[\gA]$, and
collect the terms of each colour (trace) order of $\mathcal{O}(\gB^4)$.
Contracting external lines in a colour-ordered manner into these terms, we
have
\begin{equation}
  \begin{split}
    V^2(1234) = \phantom{+} & \frac{\hat 1}{\hat 5} V^2(5^A23) \Xi^2(\bar
    5^A41)
    +   \frac{\hat 2}{\hat 5} V^2(15^A4) \Xi^1(\bar 5^A23) \\
    + & V^2(125^A) \Upsilon(\bar 5^A34) + W^2(1234)
  \end{split}
  \label{eq:vtx-4-2}
\end{equation}
for the $\bar\gB_I(1)\bar\gB_J(2)\gB_K(3)\gB_L(4)$ vertex, and
\begin{equation}
  \begin{split}
    V^3(1234) = \phantom{+} & \frac{\hat 1}{\hat 5} V^2(5^A34) \Xi^1(\bar
    5^A12)
    +   \frac{\hat 3}{\hat 5} V^2(15^A4) \Xi^2(\bar 5^A23) \\
    + & \frac{\hat 3}{\hat 5} V^2(5^A12) \Xi^2(\bar 5^A34)
    +   \frac{\hat 1}{\hat 5} V^2(35^A2) \Xi^1(\bar 5^A41) \\
    + & 2W^3(1234)
  \end{split}
  \label{eq:vtx-4-3}
\end{equation}
for the $\bar\gB_I(1)\gB_J(2)\bar\gB_K(3)\gB_L(4)$ vertex.

That these expressions reduce in four dimensions should be obvious by
comparing the forms of \eqref{eq:vtx-4-2} and \eqref{eq:vtx-4-3} to their
four-dimensional analogs in ref.  \cite{Ettle:2006bw} and noting the
reduction of the individual factors.

\section{The One-Loop $(\fourplus)$ Amplitude}
\label{sec:pppp-4cut}

It is not possible to construct a one-loop $(\fourplus)$ amplitude
using only the $\gB$ vertices of the canonical MHV lagrangian.
Nevertheless, we know it is non-vanishing (see, \eg,
\cite{Bern:1993mq, Bern:1993sx,
  Mahlon:1993si, Bern:1993qk, Bern:1995db}). We will see that it arises
  (as it indeed must)
from equivalence theorem evading pieces, constructed from the
MHV completion vertices of fig.~\ref{fig:feyn-UpsilonXi}.

In all, we can construct four classes of graphs for this
contribution: boxes, triangles, two classes of bubbles
(corresponding to the two possible arrangements of external lines
on either side of the loop), and the tadpoles.

In the next three subsections, we will consider the generalised
quadruple cut of these diagrams. We will restrict ourselves to
analysing the cuts that arise from the singularities provided by
the propagators, which we refer to as \emph{standard} cuts.
From general considerations \cite{analytics} we expect other
\emph{non-standard} cuts arising from the singular denominators in
the vertices. This is true both of the $D$ dimensional version we
have here and the four dimensional Parke-Taylor forms
\eqref{eq:pta}. From the earlier derivation it is clear that this
singular behaviour is restricted to the quantisation surface (they
have no dependence on $\check p$). These cuts therefore depend on
the orientation of the quantisation surface, \ie\ $\mu$, and are
thus gauge artifacts which should all cancel out in any complete
on-shell amplitude.

In general, by the Feynman tree theorem \cite{Brandhuber:2005kd,
Feynman:1972mt}, we can reconstruct the amplitude entirely using only the
standard cuts since only these involve a change in light cone time. However,
we will not follow this procedure here since we will be content to show
instead in the last two subsections that the diagrams we construct simply
sum up to the contributions one would have obtained from the original light
cone Yang-Mills lagrangian \eqref{eq:ym-lc4d-action} (even before
integration over the loop momentum).

\subsection{Off-shell quadruple cut}
\label{ssec:pppp-box}

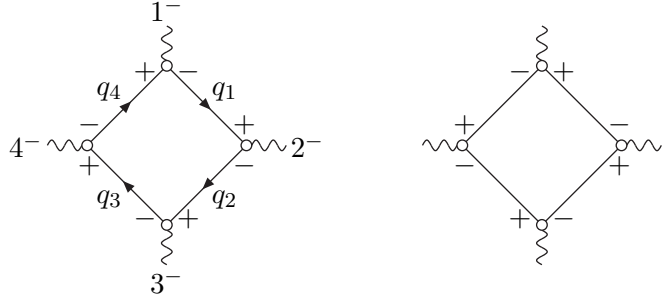
\begin{figure}[h]
  \centering \subfigure{
    \begin{picture}(120,120) \SetOffset(60,60) \Photon(0,32)(0,45){2}{2}
      \Photon(0,-32)(0,-45){2}{2} \Photon(32,0)(45,0){2}{2}
      \Photon(-32,0)(-45,0){2}{2} \ArrowLine(0,30)(30,0)
      \ArrowLine(0,-30)(-30,0) \ArrowLine(30,0)(0,-30)
      \ArrowLine(-30,0)(0,30) \BCirc(0,30){2} \BCirc(30,0){2}
      \BCirc(0,-30){2} \BCirc(-30,0){2} \Text(0,47)[bc]{$1^-$}
      \Text(47,0)[lc]{$2^-$} \Text(0,-47)[tc]{$3^-$} \Text(-47,0)[rc]{$4^-$}
      \Text(17,17)[bl]{$q_1$} \Text(17,-17)[tl]{$q_2$}
      \Text(-17,-17)[tr]{$q_3$} \Text(-17,17)[br]{$q_4$}
      \Text(4,28)[cl]{$-$} \Text(-4,28)[cr]{$+$} \Text(4,-28)[cl]{$+$}
      \Text(-4,-28)[cr]{$-$} \Text(25,4)[bl]{$+$} \Text(25,-4)[tl]{$-$}
      \Text(-25,4)[br]{$-$} \Text(-25,-4)[tr]{$+$}
    \end{picture}
    \label{feyn:mhv-box-clock}
  }\quad \subfigure{
    \begin{picture}(100,80) \SetOffset(60,60) \Photon(0,32)(0,45){2}{2}
      \Photon(0,-32)(0,-45){2}{2} \Photon(32,0)(45,0){2}{2}
      \Photon(-32,0)(-45,0){2}{2} \Line(0,30)(30,0) \Line(0,-30)(-30,0)
      \Line(30,0)(0,-30) \Line(-30,0)(0,30) \BCirc(0,30){2} \BCirc(30,0){2}
      \BCirc(0,-30){2} \BCirc(-30,0){2} \Text(4,28)[cl]{$+$}
      \Text(-4,28)[cr]{$-$} \Text(4,-28)[cl]{$-$} \Text(-4,-28)[cr]{$+$}
      \Text(25,4)[bl]{$-$} \Text(25,-4)[tl]{$+$} \Text(-25,4)[br]{$+$}
      \Text(-25,-4)[tr]{$-$}
    \end{picture}
    \label{feyn:mhv-box-anticlock}
  }
  \caption{Box contributions to the one-loop $(\fourplus)$ amplitude.  All
    external momenta are taken as outgoing. }
  \label{feyn:mhv-box}
\end{figure}

First, consider the contribution to the standard quadruple
cut coming from the box diagram, shown in fig.~\ref{feyn:mhv-box}.
The amplitude is obtained by amputating $\langle \bar A \bar A \bar A \bar A
\rangle$, the box diagram for which is shown in fig.~\ref{feyn:mhv-box}.
This gives a contribution of

\begin{equation}
\begin{split}
\label{eq:boxintegral}
A^{\text{box}}(1^+,2^+,3^+,4^+) = &\lim_{p_1^2, p_2^2, p_3^2, p_4^2
    \rightarrow 0}
    \frac 14 g^4 \:
    \frac{p_1^2 p_2^2 p_3^2 p_4^2}{\hat 1 \hat 2 \hat 3 \hat 4}
    \int \frac{d^D q}{(2\pi)^D}
    \frac{16}{q_1^2 q_2^2 q_3^2 q_4^2} 
    \frac{1}{\Sigma_1 \Sigma_2 \Sigma_3 \Sigma_4}
    \times \\
    \bigl\{ &
        \bar V^2(-q_4^D, 1, q_1^A) \bar V^2(-q_1^A, 2, q_2^B)
        \bar V^2(-q_2^B, 3, q_3^C) \bar V^2(-q_3^C, 4, q_4^D) \\
    & + \bar V^2(1, q_1^A, -q_4^D) \bar V^2(2, q_2^B, -q_1^A)
        \bar V^2(3, q_3^C, -q_2^B) \bar V^2(4, q_4^D, -q_3^C)
    \bigr\},
\end{split}
\end{equation}
where we have already used \eqref{eq:xi-3pt}, the internal
momenta are defined as $q_i=q-P_{1i}$, and we define the short-hand
\begin{equation}
\label{eq:Sigma-sh}
\Sigma_j := \frac{q_j^2}{\hat q_j} - \frac{q_{j-1}^2}{\hat q_{j-1}}
    + \frac{p_j^2}{\hat p_j}
\end{equation}
(indices interpreted cyclically). Note that the external momenta $p_1,
\dots, p_4$ are in four dimensions and thus their transverse indices have
all been set to one.

Before going on to compute the quadruple standard cut of the box, as an
aside we will show that its double and triple standard cuts vanish for
on-shell external momenta. Consider cutting any three internal lines of
fig.~\ref{feyn:mhv-box}, where by this we mean strictly to analyse only
those parts of phase space where the remaining internal line stays
off-shell.  Without loss of generality, we choose these internal lines to be
$q_1$, $q_2$ and $q_3$.  In order that the amplitude survives LSZ reduction,
the correlator must generate a singularity in $p_1^2 p_2^2 p_3^2 p_4^2$.
Clearly, the $\Sigma_2$ and $\Sigma_3$ denominators provide a singularity
$p_2^2 p_3^2$ once the internal lines are cut. We now claim that the triple
cut vanishes as follows: the required singularity in $p_1^2p_4^2$ must come
from the denominators in the remaining tree graph connecting $p_1$ and
$p_4$. The relevant factors from \eqref{eq:boxintegral} are the $q_4$
propagator, and the $\Sigma_1$ and $\Sigma_4$ denominators, \ie\
\[
\frac1{q_4^2}
\left(\frac{q_1^2}{\hat q_1} - \frac{q_4^2}{\hat q_4}
    + \frac{p_1^2}{\hat p_1}\right)^{-1}
\left(\frac{q_4^2}{\hat q_4} - \frac{q_3^2}{\hat q_3}
    + \frac{p_4^2}{\hat p_4}\right)^{-1}.
\]
Upon setting $q_1^2$ and $q_3^2$ to zero and discarding (the non-vanishing)
factors of momenta that appear on the numerator, we arrive at
\[
\frac1{q_4^2}
\frac 1{\hat q_4\,p_1^2 - \hat 1  \,q_4^2} \frac 1{\hat 4  \,q_4^2
+ \hat q_4\,p_4^2}.
\]
The above factors clearly cannot cancel $p_1^2 p_4^2$ so long as $q_4^2 \ne
0$. Hence this cut vanishes as we take all the external momenta on shell. By
similar consideration, one can also see that both possible double cuts of
this graph also vanish.

Now, we will compute the standard quadruple cut. This is obtained by putting
all four internal lines on shell \cite{analytics, Britto:2004nc,
Brandhuber:2005jw}.  The external momenta are kept off shell momentarily. We
see that the $\Sigma_i$ reduce to $p_i^2/\hat i$ factors upon cutting,
producing poles which thus cancel the factors of $p^2_i$ from LSZ reduction
and the $1/\hat i$ factors. The remaining terms have a finite
non-vanishing\footnote{Recall that four-cut solutions are non-vanishing
because they use complex external and internal momenta \cite{analytics,
Britto:2004nc, Brandhuber:2005jw}.} on-shell limit and it is already clear
that they are exactly what we obtain from the four-cut box contribution
using the light cone Yang-Mills lagrangian \eqref{eq:ym-lc4d-action}.

For the purposes of demonstrating that precisely this contribution arises from
the four-cut MHV completion box graph within the present formalism we do not
need to go any further. However, let us show how this contribution can be
straightforwardly computed within the bracket formalism we have developed here
and in ref. \cite{Ettle:2006bw}.

\subsection{Explicit evaluation}

First note that the external momenta remain four-dimensional, whereas the
loop momentum $q \equiv q_4$ in the integral is $D$-dimensional. The
solution to the four constraints $q^2_i=0$ fixes the four-dimensional part
of $q$ to a discrete set of solutions (in fact two); these are functions of
the remaining, orthogonal $-2\epsilon$ components $\mu$. Now since $q$ can
only contract with either itself or the four-dimensional external momenta,
we see that these solutions can in fact only depend on $\mu^2 = 2(q_I \bar
q_I - \tilde q \bar q)$. 

All external momenta are now on shell in the forthcoming analysis.
Substituting for the vertices, we find the remaining cut integral
to be
\begin{equation}
\label{eq:cutintegral}
8 g^4 (1-\epsilon) \frac{1}{\hat 1 \hat 2 \hat 3 \hat 4}
    \int \frac{d\mu^{-2\epsilon}}{(2\pi)^{-2\epsilon}}
    \{q\:1\} \{q\!-\!1,2\} \{q\!+\!4,3\} \{q\:4\},
\end{equation}
where we have split the integral over momentum space as in
\cite{Bern:1995db}.  Note that the $\left\{ \cdots \right\}$ bilinears above
have their index set to $1$, but this has been dropped for clarity.

We also
remind the reader that the four dimensional part of $q$ is now a function of
$\mu^2$, and note that the factor of $(1-\epsilon)$ comes from dimensional
regularisation of the gauge field degrees of freedom (as opposed to
`dimensional reduction', which only extends the loop momentum to $D$
dimensions and therefore lacks this factor).
In the following, since the momenta are
complex, $\{q\:1\}$ is not related by complex conjugation to
$(q\:1)$. We also
remind the reader that the four dimensional part
Solutions for the bilinears in \eqref{eq:cutintegral} are evaluated
directly.  First, consider $(q\:1) \mdot \{q\:1\}$.  Since, $q_1^2 = q^2 =
p^2_1 = 0$, \eqref{eq:null-p-dot-q}, or \eqref{eq:sum-bilinears}, implies
that this vanishes. Splitting away the four-dimensional part gives
\begin{align}
\label{eq:bilinear-mu-1}
(q\:1)\{q\:1\} + \hat 1^2 \mu^2/2 &= 0, \\
\intertext{and similarly}
\label{eq:bilinear-mu-2}
(q\!-\!1,2)\{q\!-\!1,2\} + \hat 2^2 \mu^2/2 &= 0 \quad\text{and}\quad \\
\label{eq:bilinear-mu-3}
(q\:4)\{q\:4\} + \hat 4^2 \mu^2/2 &= 0.
\end{align}
We eliminate $\mu^2$ between \eqref{eq:bilinear-mu-1},
\eqref{eq:bilinear-mu-2} and \eqref{eq:bilinear-mu-3}, and then use
\eqref{eq:bianchi} to eliminate $(q\:4)$ and its conjugate to obtain a quadratic
equation
\begin{equation}
\label{eq:q1-quadratic}
\alpha (q\:1)^2 + \hat 1 (q\:1) - \bar\alpha \frac{\hat 1^2 \mu^2}{2} = 0
\end{equation}
where\footnote{Again, note that here $\bar\alpha\ne\alpha^*$.}
\begin{equation}
\label{eq:bilinear-alphas}
    \alpha = \frac{\hat 4}{ (1\:4) } - \frac{\hat 2}{ (1\:2) },\quad
\bar\alpha = \frac{\hat 4}{\{1\:4\}} - \frac{\hat 2}{\{1\:2\}}.
\end{equation}
This has solutions
\begin{equation}
\label{eq:q1-solutions}
  (q\:1) = -\frac{\hat 1}{2\alpha} (1 \pm \beta),\quad
\{q\:1\} = -\frac{\hat 1}{2\bar\alpha} (1 \mp \beta),\quad
   \beta = \sqrt{1 + 2\alpha\bar\alpha \mu^2}.
\end{equation}
Next, the Bianchi-like identity \eqref{eq:bianchi} gives
\begin{equation}
\hat 1\{q\!-\!1,2\} = \hat 2\{q\:1\} + (\hat q - \hat 1)\{1\:2\}.
\label{eq:q12-bianchi}
\end{equation}
We apply this equation, its conjugate, and \eqref{eq:bilinear-mu-1} to
\eqref{eq:bilinear-mu-2} to obtain an expression for $\hat q - \hat 1$ in
terms of $(q\:1)$ and $\{q\:1\}$. Inserting this back into
\eqref{eq:q12-bianchi} and using \eqref{eq:q1-solutions} gives
\begin{equation}
\label{eq:q12-solutions}
\{q\!-\!1,2\} = \frac{\hat 2}{2\alpha}\frac{\{1\:2\}}{(1\:2)}(1 \pm \beta).
\end{equation}
Similarly, we find
\begin{equation}
\label{eq:q4-solutions}
\{q\:4\} = \frac{\hat 4}{2\alpha}\frac{\{1\:4\}}{(1\:4)}(1 \pm \beta).
\end{equation}
To obtain the final bilinear, we use \eqref{eq:bianchi} twice to obtain
$\{q\!+\!4, 3\}$ in terms of $\{q\:1\}$ and $\{q\:4\}$.
\eqref{eq:sum-bilinears} is then applied to eliminate a quotient of
$(\cdots)$ bilinears present in one of the terms in favour of conjugate
bilinears, giving
\begin{equation}
\label{eq:q43-solutions}
\{q\!+\!4,3\} = \frac 12
    \frac{\{2\:3\}\{3\:4\}}{\{2\:4\}} (1 \pm \beta).
\end{equation}

Assembling the product of the $\{\cdots\}$ bilinears from
\eqref{eq:q1-solutions}, \eqref{eq:q12-solutions}, \eqref{eq:q4-solutions} and
\eqref{eq:q43-solutions}, we have
\begin{equation}
\{q\:1\} \{q\!-\!1,2\} \{q\!+\!4,3\} \{q\:4\}
    = - \tfrac 14 \hat 1^2 \hat 2 \hat 4 \:\mu^4\:
        \frac{\{2\:3\}\{3\:4\}}{(1\:2)(4\:1)}
    = - \tfrac 14 \hat 1 \hat 2 \hat 3 \hat 4 \:\mu^4\:
        \frac{\{1\:2\}\{3\:4\}}{(1\:2)(3\:4)}
\end{equation}
for either of the solutions \eqref{eq:q1-solutions},
where in the second assertion we have used the fact that the right-hand side of
\eqref{eq:sum-bilinears} is zero for null $p_j$. Using this in
\eqref{eq:cutintegral} and reinstating the propagators, we arrive at
\begin{equation}
\label{eq:cutintegral-lifted}
2(1-\epsilon) g^4 \frac{\{1\:2\}\{3\:4\}}{(1\:2)(3\:4)}
    \int \frac{d^4q\:d^{-2\epsilon}\mu}{(2\pi)^D}
         \frac{\mu^4}{q_1^2 q_2^2 q_3^2 q_4^2}
\end{equation}
from which one observes that \eqref{eq:boxintegral} has precisely the quadruple
cut of the $4-2\epsilon$-dimensional box function $K_4$, as expected of this
amplitude \cite{Bern:1993mq, Bern:1993sx, Mahlon:1993si, Bern:1993qk,
Bern:1995db}.

\subsection{Triangle, bubble and tadpole contributions}
\label{ssec:pppp-trianglebubble}

Typical triangle, bubble and tadpole contributions to the one-loop
$(\fourplus)$ amplitude with internal helicities running from $-$ to $+$ in
a clockwise sense\footnote{The ``sense'' of internal helicity orientation is
always defined in this paper as propagating from $-$ to $+$.} are shown in
figs.~\ref{fig:mhv-triangles}, \ref{fig:mhv-bubbles} and
\ref{fig:mhv-tadpoles}.
\begin{figure}[h]
  \centering \subfigure{
    \begin{picture}(120,120)(0,0) \ArrowLine(90,30)(30,90)
      \ArrowLine(90,90)(90,30) \ArrowLine(30,90)(90,90)
      \ArrowLine(30,90)(30,30) \Photon(10,10)(30,30){2}{3}
      \Photon(110,10)(90,30){2}{3} \Photon(110,110)(90,90){2}{3}
      \Photon(10,110)(30,90){2}{3} \BCirc(30,30){2} \BCirc(90,30){2}
      \BCirc(90,90){2} \BCirc(30,90){2} \Text(15,10)[tr]{$4^-$}
      \Text(110,10)[tl]{$3^-$} \Text(15,110)[br]{$1^-$}
      \Text(103,111)[bl]{$2^-$} \Text(21,60)[cc]{$p_4$}
      \Text(60,100)[cc]{$q_1$} \Text(100,60)[cc]{$q_2$}
      \Text(58,50)[cc]{$q_3$} \Text(22,35)[cc]{$-$} \Text(22,83)[cc]{$+$}
      \Text(38,97)[cc]{$-$} \Text(82,97)[cc]{$+$} \Text(98,35)[cc]{$+$}
      \Text(98,83)[cc]{$-$} \Text(80,30)[cc]{$-$} \Text(37,75)[cc]{$+$}
    \end{picture}
  } 
  \quad
  \subfigure{
    \begin{picture}(120,120)(0,0) \ArrowLine(90,30)(30,30)
      \ArrowLine(90,90)(90,30) \ArrowLine(30,30)(90,90)
      \ArrowLine(30,30)(30,90) \Photon(10,10)(30,30){2}{3}
      \Photon(110,10)(90,30){2}{3} \Photon(110,110)(90,90){2}{3}
      \Photon(10,110)(30,90){2}{3} \BCirc(30,30){2} \BCirc(90,30){2}
      \BCirc(90,90){2} \BCirc(30,90){2} \Text(15,10)[tr]{$4^-$}
      \Text(110,10)[tl]{$3^-$} \Text(15,110)[br]{$1^-$}
      \Text(103,111)[bl]{$2^-$} \Text(21,60)[cc]{$p_1$}
      \Text(60,20)[cc]{$q_3$} \Text(100,60)[cc]{$q_2$}
      \Text(60,65)[br]{$q_1$} \Text(22,35)[cc]{$+$} \Text(22,83)[cc]{$-$}
      \Text(38,45)[cc]{$-$} \Text(82,90)[cc]{$+$} \Text(98,35)[cc]{$+$}
      \Text(98,83)[cc]{$-$} \Text(80,24)[cc]{$-$} \Text(37,24)[cc]{$+$}
    \end{picture}
  } 
  \caption{One-loop MHV completion triangle graphs for the $(\fourplus)$
  amplitude. Note
  that the propagator carrying an external momentum is attached to the $\Xi$
  vertex differently in each case.}
  \label{fig:mhv-triangles}
\end{figure}
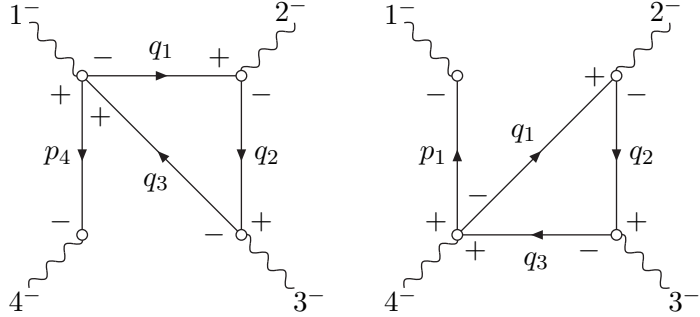
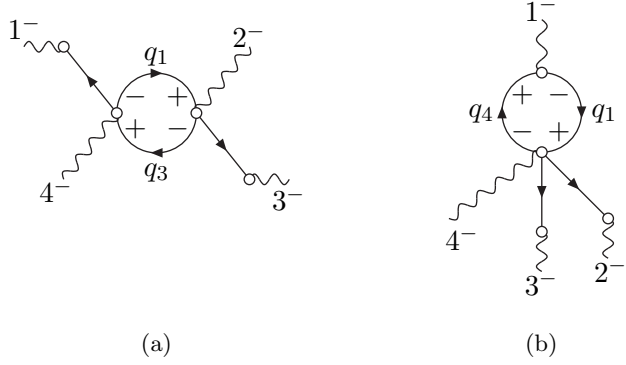
\begin{figure}
   \centering
   \subfigure[]{
    \begin{picture}(120,120)(0,0) \SetOffset(60,75)
      \ArrowLine(-15,0)(-35,25)
      \Photon(-35,-25)(-15,0){2}{4} \ArrowArcn(0,0)(15,0,180)
      \ArrowArcn(0,0)(15,180,0) \ArrowLine(15,0)(35,-25)
      \Photon(15,0)(35,25){2}{4} \Photon(-35,25)(-50,25){2}{2}
      \Photon(35,-25)(50,-25){2}{2} \BCirc(-15,0){2} \BCirc(15,0){2}
      \BCirc(-35,25){2} \BCirc(35,-25){2} \Text(35,25)[cb]{$2^-$}
      \Text(50,-28)[ct]{$3^-$} \Text(-38,-25)[ct]{$4^-$}
      \Text(-50,28)[cb]{$1^-$} \Text(0,18)[cb]{$q_1$}
      \Text(0,-19)[ct]{$q_3$} \Text(8,6)[cc]{$+$} \Text(8,-6)[cc]{$-$}
      \Text(-8,6)[cc]{$-$} \Text(-8,-6)[cc]{$+$}
    \end{picture}
    \label{fig:mhv-bubbles-22}
  }
  \quad
  \subfigure[]{
    \begin{picture}(120,120)(0,0) \SetOffset(25,0)
      \Photon(35,110)(35,90){2}{2} \ArrowArcn(35,75)(15,90,270)
      \ArrowArcn(35,75)(15,270,90) \Photon(35,60)(0,35){-2}{5}
      \ArrowLine(35,60)(35,30) \Photon(35,30)(35,15){2}{2}
      \ArrowLine(35,60)(60,35) \Photon(60,35)(60,20){2}{2} \BCirc(35,90){2}
      \BCirc(35,60){2} \BCirc(35,30){2} \BCirc(60,35){2}
      \Text(35,110)[bc]{$1^-$} \Text(55,20)[tl]{$2^-$}
      \Text(35,15)[tc]{$3^-$} \Text(5,32)[tc]{$4^-$} \Text(17,75)[cr]{$q_4$}
      \Text(54,75)[cl]{$q_1$} \Text(32,82)[cr]{$+$} \Text(46,82)[cr]{$-$}
      \Text(32,68)[cr]{$-$} \Text(46,68)[cr]{$+$}
    \end{picture}
    \label{fig:mhv-bubbles-13}
  }
  \caption{MHV completion bubble graphs. In (a) we show a the $2|2$ bubble.
  There are
  three other graphs like it (up to shifting the external momentum labels
  once by $i\rightarrow i+1$)
  obtained by swapping the external momentum propagators between
  gluons $1$ and $4$, and between gluons $2$ with $3$. The $3|1$ bubble is
  shown in (b); there are two additional graphs (up to rotations of the
  labels), in this case obtained by
  associating the wavy line attached to the five-point $\Xi$ with gluon
  $2$ or $3$ instead of $4$.}
\label{fig:mhv-bubbles}
\end{figure}
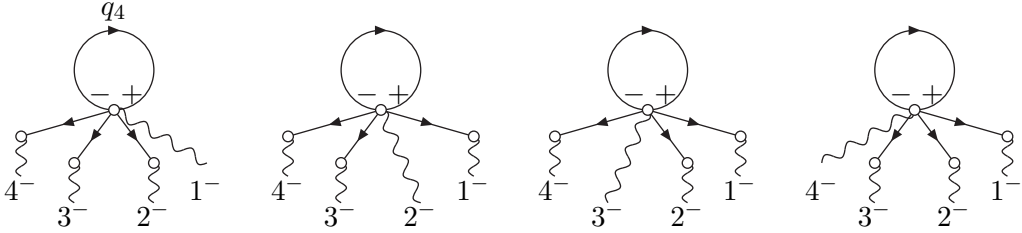
\begin{figure}[h]
  \centering
  \subfigure{
    \begin{picture}(90,87)(0,0)
      \Text(45,79)[bc]{$q_4$}
      \SetOffset(0,-5)
      \ArrowArcn(45,65)(15,269.9999,270.0001)
      \Photon(45,50)(80,30){2}{4}
      \Photon(10,40)(10,25){2}{2} \ArrowLine(45,50)(10,40) \BCirc(10,40){2}
      \Text(10,25)[tc]{$4^-$}
      \ArrowLine(45,50)(30,30)
      \Photon(30,30)(30,15){2}{2} 
      \ArrowLine(45,50)(60,30) 
      \Photon(60,30)(60,15){2}{2}  \BCirc(60,30){2}
      \BCirc(30,30){2} \BCirc(45,50){2}
       \Text(30,15)[tc]{$3^-$}
      \Text(60,15)[tc]{$2^-$} \Text(80,25)[tc]{$1^-$} \Text(40,56)[cc]{$-$}
      \Text(52,56)[cc]{$+$}
    \end{picture}
  }
  \subfigure{
    \begin{picture}(90,87)(0,0)
      \SetOffset(0,-5)
      \ArrowArcn(45,65)(15,269.9999,270.0001)
      \Photon(10,40)(10,25){2}{2} \ArrowLine(45,50)(10,40) \BCirc(10,40){2}
      \Text(10,25)[tc]{$4^-$}
      \ArrowLine(45,50)(80,40) \ArrowLine(45,50)(30,30)
      \Photon(80,40)(80,25){2}{2}
      \Photon(30,30)(30,15){2}{2} \Photon(45,50)(60,15){2}{4}
      \BCirc(80,40){2} \BCirc(30,30){2} \BCirc(45,50){2}
       \Text(30,15)[tc]{$3^-$}
      \Text(60,15)[tc]{$2^-$} \Text(80,25)[tc]{$1^-$} \Text(40,56)[cc]{$-$}
      \Text(52,56)[cc]{$+$}
    \end{picture}
  }
  \subfigure{
    \begin{picture}(90,87)(0,0)
      \SetOffset(0,-5)
      \ArrowArcn(45,65)(15,269.9999,270.0001)
      \Photon(10,40)(10,25){2}{2} \ArrowLine(45,50)(10,40) \BCirc(10,40){2}
      \Text(10,25)[tc]{$4^-$}
      \ArrowLine(45,50)(80,40) \ArrowLine(45,50)(60,30)
      \Photon(80,40)(80,25){2}{2}
      \Photon(60,30)(60,15){2}{2} \Photon(45,50)(30,15){2}{4}
      \BCirc(80,40){2} \BCirc(60,30){2} \BCirc(45,50){2}
       \Text(30,15)[tc]{$3^-$}
      \Text(60,15)[tc]{$2^-$} \Text(80,25)[tc]{$1^-$} \Text(40,56)[cc]{$-$}
      \Text(52,56)[cc]{$+$}
    \end{picture}
  }
  \subfigure{
    \begin{picture}(90,87)(0,0)
      \SetOffset(0,-5)
      \ArrowArcn(45,65)(15,269.9999,270.0001) \Photon(45,50)(10,30){2}{4}
      \Text(10,25)[tc]{$4^-$}
      \ArrowLine(45,50)(80,40) \ArrowLine(45,50)(30,30)
      \ArrowLine(45,50)(60,30) \Photon(80,40)(80,25){2}{2}
      \Photon(30,30)(30,15){2}{2} \Photon(60,30)(60,15){2}{2}
      \BCirc(80,40){2} \BCirc(30,30){2} \BCirc(60,30){2} \BCirc(45,50){2}
       \Text(30,15)[tc]{$3^-$}
      \Text(60,15)[tc]{$2^-$} \Text(80,25)[tc]{$1^-$} \Text(40,56)[cc]{$-$}
      \Text(52,56)[cc]{$+$}
    \end{picture}
  }
  \caption{Tadpole MHV completion graphs. Notice that the coupling of the
  six-point $\Xi$ to the $\bar A$ field
  (denoted by the wavy line) is associated with a different gluon in
  each case.}
  \label{fig:mhv-tadpoles} 
\end{figure}

Despite appearances, these diagrams do have quadruple cuts as a result of
the singularities in the vertices. We therefore have to consider also
cutting the vertices. Let us first consider the quadruple cut of the
triangle graph. We can restrict our analysis to the graphs
fig.~\ref{fig:mhv-triangles}. Their contribution
to the $(\fourplus)$ amplitude is
\begin{equation}
\label{eq:pppp-tri-Xi}
\begin{split}
  \lim_{p_1^2, p_2^2, p_3^2, p_4^2
    \rightarrow 0}
  \frac 14 g^4 \:
  \frac{p_1^2 p_2^2 p_3^2 p_4^2}{\hat 1 \hat 2 \hat 3 \hat 4}
  \int &\frac{d^Dq}{(2\pi)^D}
  \frac{\hat q_1 \hat q_2 \hat q_3}{q_1^2 q_2^2 q_3^2} \times \\
  \biggl\{ & - \Xi^1(1, q_1^A, -q_3^C, 4) \Xi^1(2, q_2^B, -q_1^A)
  \Xi^1(3, q_3^C, -q_2^B) \frac{\hat 4}{p_4^2} \\
  & - \Xi^2(4, 1, q_1^A, -q_3^C) \Xi^2(2, q_2^B, -q_1^A)
  \Xi^2(3, q_3^C, -q_2^B) \frac{\hat 1}{p_1^2} \biggr\}.
\end{split}
\end{equation}
Using the recursion relations \eqref{eq:upsilon-rr} and \eqref{eq:xi-rr}, we
can re-write this as
\begin{equation}
\label{eq:pppp-tri-Vbar2}
\begin{split}
  \lim_{p_1^2, p_2^2, p_3^2, p_4^2
    \rightarrow 0}
  \frac 14 g^4 \:
  \frac{p_1^2 p_2^2 p_3^2 p_4^2}{\hat 1 \hat 2 \hat 3 \hat 4}
  \int \frac{d^Dq}{(2\pi)^D} \frac{16}{q_1^2 q_2^2 q_3^2} \Bigg\{ 
  \frac{X}{\hat q_4 \Sigma_1 \Sigma_2 \Sigma_3 \Sigma_4 (\Sigma_1 +
  \Sigma_4)} \left(
    \frac{\Sigma_1 \hat 4}{p_4^2}- \frac{\Sigma_4 \hat 1}{p_1^2}
    \right)  \\
   + \frac{Y}{(\hat 1 + \hat 4) \Sigma'_{1+4} \Sigma_2 \Sigma_3} 
  \left(  \frac 1 {\Sigma_{1+4}} - \frac{1}{\Sigma_1 + \Sigma_4}
  \right) \left(  \frac{\hat 1}{p_1^2} + \frac{\hat 4}{p_4^2}
  \right)   \Bigg\}
\end{split}
\end{equation}
with
\begin{align}
X =&   \bar V^2(-q_4^D, 1, q_1^A) \bar V^2(-q_1^A, 2, q_2^B)
       \bar V^2(-q_2^B, 3, q_3^C) \bar V^2(-q_3^C, 4, q_4^D),
\label{eq:pppp-tri-X} \\
Y =&   \bar V^2(4, 1, 2+3^D)  \bar V^2(-q_3^C, 1+4^D, q_1^A)
       \bar V^2(-q_1^A, 2, q_2^B) \bar V^2(-q_2^B, 3, q_3^C),
\label{eq:pppp-tri-Y}
\end{align}
$q_4 := q_3 - p_4$ flowing ``through'' the vertex attached to $p_1$ (or
$p_4$) as
part of a box-like momentum-flow topology (see section \ref{ssec:pppp-lc-recon}
below for a further investigation of this), and we define the following
extensions of $\Sigma_i$ as
\begin{align}
 \Sigma_{1+4} &:= \frac{q_1^2}{\hat q_1} - \frac{q_3^2}{\hat q_3}
    + \frac{(p_1 + p_4)^2}{\hat 1 + \hat 4}, \\
\Sigma'_{1+4} &:= \frac{p_1^2}{\hat 1} + \frac{p_4^2}{\hat 4}
    - \frac{(p_1 + p_4)^2}{\hat 1 + \hat 4}.
\end{align}
Note that one can write down expressions for the analogues of $X$ and $Y$
from graphs with internal helicities of an anti-clockwise sense. The reader
may check for the case at hand (where $\bar V^2$ is the three-point \mhvbar\
vertex), that these are the same as in the clockwise scenario.

Now, recall that we are only studying the standard cuts. We will extract such a
quadruple cut contribution here, by keeping the external momenta off the mass
shell, and look for any terms containing $1/q^2_4$ in addition to the three
propagators already appearing in \eqref{eq:pppp-tri-Vbar2}. Clearly, by
inspection of the $\Sigma_i$ factors in \eqref{eq:pppp-tri-Vbar2}, no such
$1/q^2_4$ are generated. Indeed it is impossible to generate such terms from the
vertices since the singularity in $1/q^2_4$ is not restricted to the
quantisation surface. Although the inverse $\Sigma_i$s and $\Sigma_{1+4}$ appear
in \eqref{eq:Sigma-sh} and above to yield singularities that look superficially
similar to those from propagators, by \eqref{eq:sum-omega} these terms do not
contain $\check q$ components and thus their singularities lie entirely within
the quantisation surface.

A similar analysis of the graphs of figs.~\ref{fig:mhv-bubbles} and
\ref{fig:mhv-tadpoles} leads one quickly to the same conclusion: they have
no contribution to the quadruple cut for off-shell external momenta, because
in this region none of the denominators from their $\Xi$ vertices form the
necessary propagators.\footnote{A quick way to generate
fig.~\ref{fig:mhv-bubbles}a is to note that it can be formed by replacing
the product of two three-point $\Xi$s in each term in \eqref{eq:pppp-tri-Xi}
with a single four-point $\Xi$ having the relevant arguments.} Hence, we see
that for the one-loop $(\fourplus)$ amplitude in the canonical MHV
lagrangian formalism, if we keep the external momenta off shell until after
the cuts, only the box graph of fig.~\ref{feyn:mhv-box} contributes to
the quadruple cut.

\subsection{Light-cone Yang-Mills reconstructions}
\label{ssec:pppp-lc-recon}

The expressions \eqref{eq:boxintegral}, \eqref{eq:pppp-tri-X} and
\eqref{eq:pppp-tri-Y} begin to elucidate the underlying relationship between
the MHV completion graphs of
figs.~\ref{feyn:mhv-box}--\ref{fig:mhv-tadpoles} and the Feynman graphs one
would use to compute the same amplitude in conventional perturbative LCYM.
We already see parallels of the latter in the topology of the linking
amongst $\bar V^2$ \mhvbar\ vertices.\footnote{Recall that these are the
same as the $\ppm$ vertices in light-cone Yang-Mills.}

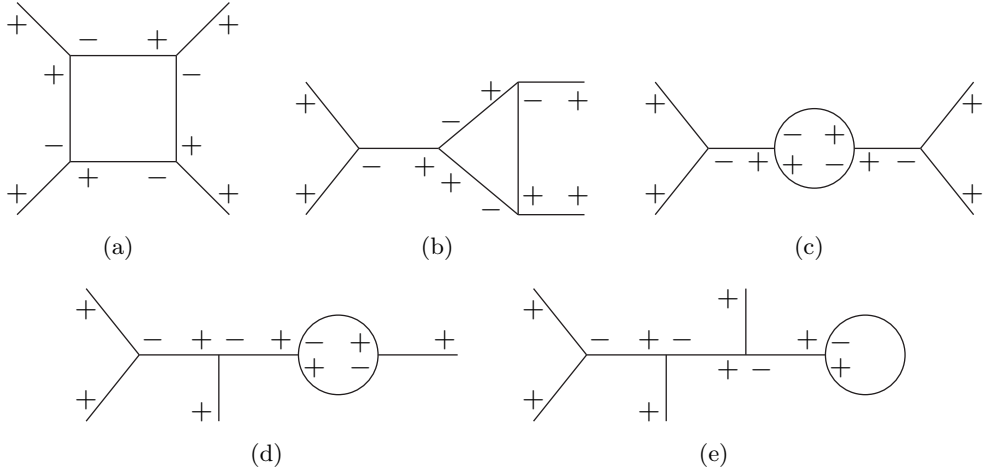
\begin{figure}[h]
\centering
\subfigure[]{
  \begin{picture}(80,80)
    \Line(0,0)(20,20) \Line(80,0)(60,20) \Line(20,20)(60,20)
    \Line(0,80)(20,60) \Line(80,80)(60,60) \Line(20,60)(60,60)
    \Line(20,20)(20,60) \Line(60,20)(60,60)
    \Text(0,8)[cc]{$+$} \Text(0,72)[cc]{$+$}
    \Text(80,72)[cc]{$+$} \Text(80,8)[cc]{$+$}
    \Text(14,26)[cc]{$-$} \Text(66,26)[cc]{$+$}
    \Text(14,53)[cc]{$+$} \Text(66,53)[cc]{$-$}
    \Text(27,14)[cc]{$+$} \Text(53,14)[cc]{$-$}
    \Text(27,66)[cc]{$-$} \Text(53,66)[cc]{$+$}
  \end{picture}
}\qquad
\subfigure[]{
  \begin{picture}(103,50)
    \Line(0,0)(20,25) \Line(0,50)(20,25)
    \Line(20,25)(50,25) 
    \Line(50,25)(80,0) \Line(50,25)(80,50)
    \Line(80,0)(80,50)
    \Line(80,0)(105,0) \Line(80,50)(105,50)
    \Text(0,8)[cc]{$+$} \Text(0,42)[cc]{$+$}
    \Text(103,7)[cc]{$+$} \Text(103,43)[cc]{$+$}
    \Text(25,18)[cc]{$-$} \Text(45,18)[cc]{$+$}
    \Text(55,35)[cc]{$-$} \Text(55,12)[cc]{$+$}
    \Text(70,47)[cc]{$+$} \Text(70,2)[cc]{$-$}
    \Text(86,7)[cc]{$+$} \Text(86,43)[cc]{$-$}
  \end{picture}
}\qquad
\subfigure[]{
  \begin{picture}(120,50)
    \Line(0,0)(20,25) \Line(0,50)(20,25)
    \Line(20,25)(100,25)
    \BCirc(60,25){15}
    \Line(100,25)(120,0) \Line(100,25)(120,50)
    \Text(0,8)[cc]{$+$} \Text(0,42)[cc]{$+$}
    \Text(120,8)[cc]{$+$} \Text(120,42)[cc]{$+$}
    \Text(26,20)[cc]{$-$} \Text(39,20)[cc]{$+$}
    \Text(81,20)[cc]{$+$} \Text(95,20)[cc]{$-$}
    \Text(52,30)[cc]{$-$} \Text(68,30)[cc]{$+$}
    \Text(52,19)[cc]{$+$} \Text(68,19)[cc]{$-$}
  \end{picture}
}\\
\subfigure[]{
  \begin{picture}(140,50)
    \Line(0,0)(20,25)
    \Line(0,50)(20,25)
    \Line(20,25)(110,25)
    \Line(50,0)(50,25)
    \Line(110,25)(140,25)
    \BCirc(95,25){15}
    \Text(0,8)[cc]{$+$}
    \Text(0,42)[cc]{$+$}
    \Text(48,0)[br]{$+$}
    \Text(78,27)[br]{$+$}
    \Text(21,27)[bl]{$-$}
    \Text(48,27)[br]{$+$}
    \Text(52,27)[bl]{$-$}
    \Text(140,27)[br]{$+$}
    \Text(82,26)[bl]{$-$}
    \Text(82,24)[tl]{$+$}
    \Text(108,26)[br]{$+$}
    \Text(108,24)[tr]{$-$}
  \end{picture}
}\qquad
\subfigure[]{
  \begin{picture}(140,50)
    \Line(0,0)(20,25)
    \Line(0,50)(20,25)
    \Line(20,25)(110,25)
    \Line(50,0)(50,25)
    \Line(80,25)(80,50)
    \BCirc(125,25){15}
    \Text(0,8)[cc]{$+$}
    \Text(0,42)[cc]{$+$}
    \Text(48,0)[br]{$+$}
    \Text(78,50)[tr]{$+$}
    \Text(21,27)[bl]{$-$}  
    \Text(48,27)[br]{$+$}
    \Text(52,27)[bl]{$-$}
    \Text(78,23)[tr]{$+$}
    \Text(82,23)[tl]{$-$}
    \Text(108,27)[br]{$+$}
    \Text(112,26)[bl]{$-$}
    \Text(112,24)[tl]{$+$}
  \end{picture}
}
\caption{Typical LCYM Feynman graphs that contribute to the $(\fourplus)$
amplitude. Shown topologies are the (a) box, (b) triangle,
  (c) bubble, (d) typical external leg correction, and (e) one of the
  tadpoles (there is another tadpole not shown here with the loop attached
  to the central leg of the tree instead).}
\label{fig:pppp-lc-box-tri-bub}
\end{figure}

Starting with expression \eqref{eq:boxintegral} for the box graph,
it is immediately apparent that the momentum routing through the
$\bar V^2$s yields the two box-like topologies in LCYM: one with the
internal helicities arranged in a clockwise sense (from the first term in
the curly braces, shown in fig.~\ref{fig:pppp-lc-box-tri-bub}a), and another
with an anticlockwise arrangement.

The triangle MHV completion graphs of fig.~\ref{fig:mhv-triangles} reveal a
mixture of topologies in the momentum routing. Naturally, one would expect
the triangle diagram of fig.~\ref{fig:pppp-lc-box-tri-bub}b, and indeed this
arises from the factor $Y$ in \eqref{eq:pppp-tri-Y}. The MHV completion
triangle graph also has terms with factors of $X$, which we know is nothing
but one of the vertex configurations in \eqref{eq:boxintegral} of box
topology, \ie\ fig.~\ref{fig:pppp-lc-box-tri-bub}a. (As we have seen
however, in this case the fourth propagator is missing.)

The bubble and tadpole graphs can be processed in a similar manner: the
graph of fig.~\ref{fig:mhv-bubbles}a contains the topology of, and therefore
contributes to the reconstruction of, the LCYM self-energy correction graph
in fig.~\ref{fig:pppp-lc-box-tri-bub}c; similarly
fig.~\ref{fig:mhv-bubbles}b contributes to the reconstruction of the
external leg corrections (an example of which is seen in
fig.~\ref{fig:pppp-lc-box-tri-bub}d). The MHV completion tadpole graphs in
fig.~\ref{fig:mhv-tadpoles} contain terms of topology of the LCYM tadpoles
of fig.~\ref{fig:pppp-lc-box-tri-bub}e (these are ill-defined but we can
take them to vanish in dimensional regularisation just as we would for the
LCYM tadpole), as well as contributing pieces with the self-energy and
external leg correction topologies. Additionally, both MHV completion
bubbles and tadpoles contribute to the reconstruction of box and triangle
LCYM graphs.

\subsection{Full reconstruction of the light-cone Yang-Mills box
contribution}

Although none of the integrands we have been discussing up to now are
identical to the contributions one obtains directly from light-cone
Yang-Mills (as a result of $\Sigma_i$ factors and in all cases except
\eqref{eq:boxintegral}, missing propagators) we now show that these
\emph{are} exactly reproduced when all the contributions of the correct
LCYM topology are summed together.

Rather than do this for all graphs displayed in
fig.~\ref{fig:pppp-lc-box-tri-bub}, we will concentrate on the diagrams of
box topology with the internal helicity configuration as displayed in
fig.~\ref{fig:pppp-lc-box-tri-bub}a. This corresponds to terms containing
the factor $X$ in \eqref{eq:pppp-tri-X}.

We have already seen that the MHV completion box graph
fig.~\ref{feyn:mhv-box} provides, through
\eqref{eq:boxintegral}, the term
\begin{equation}
  \label{eq:mhvbox-X}
  4 g^4 \:
  \int \frac{d^D q}{(2\pi)^D}
  \frac{XC}{q_1^2 q_2^2 q_3^2 q_4^2}
\end{equation}
which is identical to the LCYM box contribution,
save for the factor $C$ which is
\begin{equation}\label{eq:boxbox}
C_{\rm box} =  \frac{P_1P_2P_3P_4}{\Sigma_1 \Sigma_2 \Sigma_3
\Sigma_4}.
\end{equation}
Here we have introduced the short-hand $P_i\equiv p^2_i/\hat i$.
Similarly from the two triangle configurations in
fig.~\ref{fig:mhv-triangles}, we can read off the $X$ terms in
\eqref{eq:pppp-tri-Vbar2}, resulting in \eqref{eq:mhvbox-X} with
$C$ given by
\begin{equation}
\label{eq:trianglebox}
C_{\rm triangle} = \frac{P_2P_3}{\Sigma_2\Sigma_3}
\frac{Q_4}{\Sigma_4+\Sigma_1}\left(\frac{P_1}{\Sigma_4}-
\frac{P_4}{\Sigma_1}\right).
\end{equation}
Here we have introduced the short-hand $Q_i\equiv q^2_i/\hat i$.
The presence of $Q_4$ above is responsible for the missing
propagator in this contribution. As we remarked earlier, the
bubble diagrams of the form of fig.~\ref{fig:mhv-bubbles}a
follow straightforwardly by replacing the right half of the
diagram of fig.~\ref{fig:mhv-triangles} by terms
corresponding to the left hand half. Thus the four configurations
for these bubbles sum to give \eqref{eq:mhvbox-X} with
\begin{equation}\label{eq:22bubblebox}
C_{2|2} =
\frac{Q_2}{\Sigma_2+\Sigma_3}\left(\frac{P_3}{\Sigma_2}-\frac{P_2}{\Sigma_3}\right)
\frac{Q_4}{\Sigma_4+\Sigma_1}\left(\frac{P_1}{\Sigma_4}-\frac{P_4}{\Sigma_1}\right).
\end{equation}
The contributions from the three configurations of bubble,
fig.~\ref{fig:mhv-bubbles}b, and tadpole,
fig.~\ref{fig:mhv-tadpoles}, are more tedious to derive.
We find
\begin{equation}\label{eq:31bubblebox}
C_{3|1} = \frac{P_1}{\Sigma_1}\frac{Q_2Q_3}{\Sigma_2}\left\{
\frac{P_2}{(\Sigma_3+\Sigma_4)\Sigma_4}
-\frac{P_2+P_3}{(\Sigma_2+\Sigma_3)\Sigma_4}
+\frac{P_2+P_3+P_4}{(\Sigma_2+\Sigma_3)(\Sigma_2+\Sigma_3+\Sigma_4)}
\right\}
\end{equation}
and 
\begin{equation}
\begin{split}
C_{\rm tadpole} = \frac{Q_1Q_2Q_3}{\Sigma_1}\Biggl\{
 \frac1{(\Sigma_1+\Sigma_2)(\Sigma_1+\Sigma_2+\Sigma_3)}
 -\frac{P_1}{(\Sigma_2+\Sigma_3+\Sigma_4)(\Sigma_3+\Sigma_4)\Sigma_4}&\\
 +\frac{P_1+P_2}{(\Sigma_1+\Sigma_2)(\Sigma_3+\Sigma_4)\Sigma_4}
-\frac{P_1+P_2+P_3}{(\Sigma_1+\Sigma_2)(\Sigma_1+\Sigma_2+\Sigma_3)\Sigma_4}
&
 \Biggr\}
\end{split}
\label{eq:tadpolebox}
\end{equation}

Now note that, although there is just the one box contribution
\eqref{eq:boxbox}, there are four triangle contributions of the form
\eqref{eq:trianglebox} obtained by cyclically permuting the labels
(corresponding to the four cyclic configurations of external momenta).
Similarly there are four cyclic permutations of \eqref{eq:31bubblebox} and
\eqref{eq:tadpolebox} and two of \eqref{eq:22bubblebox}. Adding all these
contributions together, after some tedious but straightforward algebra, one
finds that these sum up to $C=1$. In other words, precisely the light-cone
Yang-Mills box contribution is recovered by summing up all of the
MHV completion contributions of this topology.

It follows immediately that we obtain the correct on-shell limit of this
amplitude \eqref{eq:cutintegral-lifted} in $D$ dimensions. However, note
well that we have obtained precisely the standard box contribution even
before we take the on-shell limit or perform any integration. The recovery
is purely an algebraic process; the MHV completion contributions are nothing
but the missing amplitudes from LCYM, rearranged.

In sec.~\ref{sec:pppp-4cut}, we obtained the correct standard four-cut
contribution by cutting the internal momenta first and only then taking the
external momenta on shell. We see again from the contributions above, that
taking the limits in this order, only the box \eqref{eq:boxbox} makes a
contribution. However, since the equality is algebraic, the same result is
obtained in whatever order one takes the
limits\footnote{Providing of course one takes the limits in the
same order in each term.}. Indeed, if one lets the external momenta
go on shell first, then only the tadpole \eqref{eq:tadpolebox}
makes a contribution, namely
\begin{equation}
\label{eq:tadpoleonshell}
C = \frac{Q_1Q_2Q_3}{(Q_1-Q_4)(Q_2-Q_4)(Q_3-Q_4)}.
\end{equation}
Adding the three cyclic permutations to this, of course one finds,
after simplification, that $C=1$ again. The tadpole survives this
limit because the four different $\Xi^s$ from the four different
configurations in fig.~\ref{fig:mhv-tadpoles} all provide,
upon expansion with \eqref{eq:xi-rr}, the same six-point
$\Upsilon(q^A,-q^A,1,2,3,4)$. By \eqref{eq:upsilon-rr-x} this
has in the denominator a term which cancels the sum over external
inverse propagators arising from LSZ reduction:
\[
\Omega_q+\Omega_{-q}+\sum_{i=1}^4\Omega_i
=-\frac12\sum_{i=1}^4P_i
\]
(where momentum conservation and \eqref{eq:sum-omega} have been
used).

In fact it is easy to derive the contribution \eqref{eq:tadpoleonshell} in
this limit. In \eqref{eq:upsilon-rr-x}, we keep only the terms in which each
factor depends on the loop momentum $q$, since any factor independent of $q$
corresponds to a tree-level decoration and thus does not have LCYM box
topology. However, only the last term satisfies this and thus for the box
topology, or more generally the $n$-gonal topology generated by an $n$-point
tadpole, we can make the replacement
\[
\Upsilon(q^A_n,-q^A,1,\dots,n) \rightarrow -
  \frac{\bar V^2(-q^B_{n-1},n,q^A_n)
  \Upsilon(q^B_{n-1},-q^A,1,\dots,n-1)}
  {\hat q_n(\Omega_{q_n}-\Omega_q+ \sum_{i=1}^n\Omega_i)}.
\]
Iterating this starting with $n=4$, and recalling the missing
propagators, one readily obtains \eqref{eq:tadpoleonshell} and the
corresponding term in \eqref{eq:tadpolebox}.

\section{Discussion and Conclusions}
\label{sec:discussion}

We have found that amplitudes which are missing from the CSW rules
arise in this framework as a result of $S$-matrix equivalence
theorem evasion by the field transformation. We showed this
first by calculating the tree-level $(\mpp)$ amplitude, which is
non-vanishing in $(2,2)$ signature/complex momentum. The
corresponding vertex, $\bar V^2$, is the term that we eliminate
from the lagrangian in going from LCYM to one that furnishes CSW
rules.

We recovered it via \eqref{eq:feyn-ppm-raw} by computing the LSZ reduction
of the associated correlation function $\langle A \bar A \bar A\rangle$.
Since we have obtained $A$ and $\bar A$ as series in $B$ and $\bar B$, we
calculated this correlation function graphically by defining ``MHV
completion vertices'' (so-called since they allow the construction of
amplitudes that cannot be built from MHV vertices) for the series
coefficients, $\Upsilon$ and $\Xi$. This correlation function was shown to
have divergences in the external momenta, such that it was \emph{not}
annihilated when these were taken on shell. The result, even before taking
the on-shell limit, correctly reproduces the three-gluon \mhvbar\ amplitude.

The fact that this missing amplitude is reproduced in the correct
form even off shell, is no accident. As we intimated in
sec.~\ref{sec:dimreg-mhvl}, we have to recover these terms
independently of the form of the eliminated vertex $\bar V^2$.
Indeed, if we repeat the exercise in the notation of
sec.~\ref{sec:dimreg-mhvl}, the equation for the $(\ppm)$ amplitude
is simply
\[
\frac{\hat3}2\left(\sum^3_{i=1}\frac{p^2_i}{\hat
i}\right)\Upsilon(3^K1^I2^J) \left( -\frac{ig}{\sqrt{2}}\right) =
\left( -\frac{ig}{\sqrt{2}}\right) \bar V^2(1^I2^J3^K),
\]
where we have used \eqref{eq:sum-omega} and \eqref{eq:upsilon-3}.
The vertex and the amplitude are thus recovered off shell in $D$
dimensions, whatever its form.

Although we will not do so here, it is clear that we can recover in this way
all the discarded tree-level amplitudes, including those illustrated in
fig.~\ref{fig:feyn-lc-mppp}. Although these amplitudes vanish on shell, they
obviously contribute to off-shell processes such as sub-processes in Standard
Model contributions, for example attached to heavy quarks, or at finite
virtuality from attaching to hadron wavefunctions. The simple form for the
eqns. \eqref{eq:Upsilon-4d} and \eqref{eq:Xi-4d} suggests that these
off-shell amplitudes also take simple forms in this formalism, just as the
on-shell amplitudes do.

It is also the case that the amplitudes in
fig.~\ref{fig:feyn-lc-mppp} no longer vanish in $D$ dimensions,
and this is important in recovering the missing one-loop
amplitudes. Otherwise, the three-particle and two-particle cuts of
the LCYM box contribution, fig.~\ref{fig:pppp-lc-box-tri-bub}a
would vanish on shell, in contradiction with the known answer
\eqref{eq:cutintegral-lifted}.

Recall that we have seen that the algebraic equivalence between
the formalism we have presented here and the LCYM theory, extends
also to the quantum level. The exact one-loop LCYM box
contribution to the ($\fourplus$) amplitude was recovered from the
diagrams constructed from the $\Xi$ MHV completion vertices
before taking any
limits or performing the loop momentum integration. The
contributions that sum to these `lost' amplitudes are readily
recognised by expressing them in terms of $\bar V^2$ and using
these to extract the relevant topologies. Although we concentrated
on this one topology it is clear that all the topologies in
fig.~\ref{fig:pppp-lc-box-tri-bub} will be recovered in this way.

In order for the quantum corrections we have been discussing to be
well defined we need to incorporate a regularisation scheme for
MHV diagrams. We augmented the usual light-cone co-ordinates in a
manner that allows us to preserve the ideas of positive and
negative `helicity' and hence apply the canonical transform to all
the degrees of freedom outside four dimensions. The series
solution to this change of variables follows in a manner similar
to the four-dimensional case, but the results are not as simple.
In particular, we do not have all orders compact holomorphic
expressions for the series coefficients or for the generalisations
of the tree-level MHV rules. However we do gain a lagrangian
containing only MHV interactions with two negative helicities and
any number of positive helicities. As in the four dimensional
case, this has the consequence of imposing an MHV rules `grading'
on perturbation theory so that in particular at tree level and one
loop, only amplitudes with two or more negative helicities can be
created by sewing together lagrangian vertices. We inherit its
simplicity in the sense that at tree-level the first non-vanishing
amplitudes are MHV amplitudes, these coinciding with the
lagrangian vertices, that NMHV amplitudes are constructed by
joining precisely two such vertices together by the propagator and
so on.

We used this regularisation to study in detail the standard cuts of the MHV
completion box contribution (of fig.~\ref{feyn:mhv-box}) to ($\fourplus$)
amplitude, \ie\ the generalised unitarity cuts that follow from cutting the
propagators. If we leave the external momenta off shell until after the
internal propagators are cut, then we find that only its quadruple cut is
non-zero. As we noted, it is clear immediately from the expression in terms
of $\bar V^2$ that this is the quadruple cut of the LCYM box contribution.
Nevertheless, we computed it in full and confirmed that it agreed with the
known result for the full amplitude. This is supporting evidence that the
dimensional regularisation we have put in place is indeed consistent. It
also allowed us to demonstrate that the solution for the cuts can be
obtained directly in terms of the bilinear brackets
\eqref{eq:lc-d-bilinears}, which play the r\^ole of $D$ dimensional
generalisations of the familiar spinor brackets.

On the other hand the other diagrams we can make, \cf\
figs.~\ref{fig:mhv-triangles}--\ref{fig:mhv-tadpoles}, evaluated this way,
have no standard four-cut contributions. In general which diagrams
contribute depends on the way in which we take the on-shell limit. For
example, we saw that if we take the external momenta on shell first, before
cutting the diagram, then only the tadpole contributions in
fig.~\ref{fig:mhv-tadpoles} contribute. In this case these alone sum to
provide the LCYM box contribution. Since the sum total of the contributions
from all diagrams is just equal to that of LCYM, before integration, we
are free to choose how we take the on-shell limits.

Although we have not treated the higher-point all-$+$ amplitudes, we have no
doubt that they would be recovered by the methods above.

In general, by computing all the standard cuts outside four dimensions we
can reconstruct the amplitude, on or off shell, without unfolding back to
LCYM. This follows by noting that it is just these cuts of propagators that
enters the Feynman tree theorem \cite{Brandhuber:2005kd}.

Another possibility to be further explored is the direct evaluation of
integrals such as \eqref{eq:boxintegral}. So far, we have just assumed that
the integrals corresponding to individual contributions exist and either
matched their properties to known functions or recognised that summing them
gives back previously treated amplitudes. However, direct evaluation
presents a number of technical challenges, in particular providing the
correct treatment of the non-standard cuts arising from the singularity
surfaces incorporated in the $\Upsilon$ and $\Xi$ vertices.

To summarise, we can conclude that the canonical MHV lagrangian,
supplemented as required by the $\Upsilon$ and $\Xi$ coefficients as in
\eqref{eq:Wick}, merely rearranges LCYM theory: the full evaluation of any
such amplitude recovers precisely the original LCYM amplitudes. Of course it
is important here that the canonical transformation to the MHV lagrangian
has unit jacobian both in four and $D$ dimensions. The fact that we do
recover the LCYM contributions at one loop in such a straightforward way,
gives us confidence in assuming that the jacobian is anomaly free (contrary
to speculations in the literature that such an anomaly might be responsible
for the ($\fourplus$) amplitude).

Although we have concentrated on the r\^ole of the $\Upsilon$ and $\Xi$
vertices in evading the equivalence theorem, we see that they are important
in general for recovering the correct off shell structure. For example by
incorporating these, renormalisation of the canonical MHV lagrangian is as
straightforward (or indeed as problematic \cite{Leibbrandt:1983pj}) as
renormalising LCYM.

We have seen that these vertices are required for some on-shell amplitudes.
Presumably they are not required for some amplitudes where only certain
legs are off shell. We leave for the future determining precisely when we
need to utilise the $\Upsilon$ and $\Xi$ vertices in any given amplitude.

It is clear that we can perform canonical transformations to eliminate
interactions from other lagrangians, employing light cone coordinates,
generalising \eqref{eq:transform-4d}, and in so doing arrive at generalised
versions of canonical MHV lagrangians. For example we can apply this
technique to other parts of the Standard Model. Since the recovery of the
missing amplitudes by including the $\Upsilon$ and $\Xi$ vertices works
whatever the form of $\bar V^2(1^I2^J3^K)$, this mechanism will function
equally well for these generalisations (even where higher-point interactions
have been subsumed). We leave as a subject of future research the question
of whether such generalisations, and indeed the MHV framework for Yang-Mills
applied in more generality as we have been discussing, lead to sufficiently
simpler computations compared to standard methods.

\paragraph{Note added:} After this paper was completed ref.
\cite{Brandhuber:2007vm} appeared which addresses these  issues from a
different perspective.

\section*{Acknowledgements}
\label{sec:acks}

The authors would like to thank Andreas Brandhuber, Nick Evans, Doug Ross,
Bill Spence and Gabriele Travaglini for helpful comments. For financial
support, Tim and James thank the PPARC; Jonathan thanks the Richard Newitt
bursary scheme and the School of Physics and Astronomy of the University of
Southampton. Paul's research was supported in part by the 
Perimeter Institute for Theoretical Physics.

\appendix

\section{Light-cone vector identities}
\label{sec:lightcone-vectors}
The appendix gives some of the identities particular to vectors in
$D$-dimensional light-cone co-ordinates. Some of these appear in a different
form in \cite{Chakrabarti:2005ny}. First, for any two $D$-vectors $p$ and $q$,
\begin{equation}
\label{eq:dotbilinears}
(p\:q)\mdot\{p\:q\} = -\tfrac12 (\hat p\,q - \hat q\,p)^2
\end{equation}
from which it is clear that for null $p$, $q$,
\begin{equation}
\label{eq:null-p-dot-q}
(p\:q)\mdot\{p\:q\} = \hat p\,\hat q\: p\cdot q.
\end{equation}
The Bianchi-like identity
\begin{equation}
\label{eq:bianchi}
\hat i (j\:k) + \hat j (k\:i) + \hat k (i\:j) = 0
\end{equation}
holds also under replacement of the hat with any transverse component $i_I$ or
$\bar i_I$, and replacement of the bilinear with its adjoint.

For a set of momenta $\{p_j\}$ that sum to zero,
\begin{equation}
\label{eq:sum-bilinears}
\sum_j\frac{(p\:j)\mdot\{j\:q\}}{\hat j} =
    \frac{\hat p\,\hat q}{2} \sum_j \frac{p^2_j}{\hat j}
\end{equation}
for any $p$ and $q$. This is the $D$-dimensional, off-shell generalisation of
the spinor identity $\sum_j \langle p\;j \rangle [j\:q] = \langle p \rvert
(\sum_j \lvert j \rangle [ j \rvert) \lvert q ] = 0$. In four 
dimensions the above identity looks the same except that the 
dot product is simply multiplication. Also,
\begin{equation}
\label{eq:sum-omega}
\sum_j \Omega_j = -\frac 12 \sum_j \frac{p^2_j}{\hat j}.
\end{equation}
(In four dimensions the left hand side has $\omega_j$ in place of
$\Omega_j$.)



\begin{thebibliography}{88}

\bibitem{Parke:1986gb}
  S.~J.~Parke and T.~R.~Taylor,
  Phys.\ Rev.\ Lett.\  {\bf 56}, 2459 (1986).

\bibitem{Berends:1988zn}
  F.~A.~Berends and W.~T.~Giele,
  Nucl.\ Phys.\ B {\bf 313}, 595 (1989).

\bibitem{Parke:1989vn}
  S.~J.~Parke and M.~L.~Mangano,
FERMILAB-CONF-89-180-T
{\it Invited talk at the Workshop on QED Structure Functions, Ann Arbor, MI, May 22-25, 1989}

\bibitem{Witten:2003nn}
  E.~Witten,
  Commun.\ Math.\ Phys.\  {\bf 252} (2004) 189
  [arXiv:hep-th/0312171].

\bibitem{Cachazo:2004kj}
  F.~Cachazo, P.~Svrcek and E.~Witten,
  JHEP {\bf 0409}, 006 (2004)
  [arXiv:hep-th/0403047].

\bibitem{Georgiou:2004wu}
G.~Georgiou and V.~V.~Khoze,
  JHEP {\bf 0405}, 070 (2004)
  [arXiv:hep-th/0404072];

\bibitem{Wu:2004fb}
J.~B.~Wu and C.~J.~Zhu,
  JHEP {\bf 0407}, 032 (2004)
  [arXiv:hep-th/0406085];

\bibitem{Bena:2004ry}
I.~Bena, Z.~Bern and D.~A.~Kosower,
  Phys.\ Rev.\ D {\bf 71}, 045008 (2005)
  [arXiv:hep-th/0406133];

\bibitem{Wu:2004jx}
J.~B.~Wu and C.~J.~Zhu,
  JHEP {\bf 0409}, 063 (2004)
  [arXiv:hep-th/0406146];

\bibitem{Kosower:2004yz}
D.~A.~Kosower,
  Phys.\ Rev.\ D {\bf 71}, 045007 (2005)
  [arXiv:hep-th/0406175];

\bibitem{Georgiou:2004by}
G.~Georgiou, E.~W.~N.~Glover and V.~V.~Khoze,
  JHEP {\bf 0407}, 048 (2004)
  [arXiv:hep-th/0407027];

\bibitem{Britto:2005fq}
  R.~Britto, F.~Cachazo, B.~Feng and E.~Witten,
  Phys.\ Rev.\ Lett.\  {\bf 94}, 181602 (2005)
  [arXiv:hep-th/0501052].

\bibitem{Risager:2005vk}
  K.~Risager,
  JHEP {\bf 0512} (2005) 003
  [arXiv:hep-th/0508206].

\bibitem{Britto:2004ap}
R.~Britto, F.~Cachazo and B.~Feng,
  Nucl.\ Phys.\ B {\bf 715}, 499 (2005)
  [arXiv:hep-th/0412308].

\bibitem{Britto:2005dg}
R.~Britto, B.~Feng, R.~Roiban, M.~Spradlin and A.~Volovich,
  Phys.\ Rev.\ D {\bf 71}, 105017 (2005)
  [arXiv:hep-th/0503198];

\bibitem{Bern:1993qk}
  Z.~Bern, G.~Chalmers, L.~J.~Dixon and D.~A.~Kosower,
  Phys.\ Rev.\ Lett.\  {\bf 72}, 2134 (1994)
  [arXiv:hep-ph/9312333].

\bibitem{Mahlon:1993si}
  G.~Mahlon,
  Phys.\ Rev.\  D {\bf 49} (1994) 4438
  [arXiv:hep-ph/9312276].

\bibitem{Bern:2005hs}
  Z.~Bern, L.~J.~Dixon and D.~A.~Kosower,
  Phys.\ Rev.\  D {\bf 71} (2005) 105013
  [arXiv:hep-th/0501240].

\bibitem{Berger:2006vq}
  C.~F.~Berger, Z.~Bern, L.~J.~Dixon, D.~Forde and D.~A.~Kosower,
  Phys.\ Rev.\  D {\bf 75}, 016006 (2007)
  [arXiv:hep-ph/0607014].

\bibitem{Britto:2004nj}
  R.~Britto, F.~Cachazo and B.~Feng,
  Phys.\ Rev.\ D {\bf 71}, 025012 (2005)
  [arXiv:hep-th/0410179].

\bibitem{Bern:2005cq}
  Z.~Bern, L.~J.~Dixon and D.~A.~Kosower,
  Phys.\ Rev.\ D {\bf 73} (2006) 065013
  [arXiv:hep-ph/0507005].


\bibitem{Bern:1994cg}
  Z.~Bern, L.~J.~Dixon, D.~C.~Dunbar and D.~A.~Kosower,
  Nucl.\ Phys.\ B {\bf 435} (1995) 59
  [arXiv:hep-ph/9409265].

\bibitem{Brandhuber:2005jw}
  A.~Brandhuber, S.~McNamara, B.~J.~Spence and G.~Travaglini,
  JHEP {\bf 0510} (2005) 011
  [arXiv:hep-th/0506068].

\bibitem{Bern:2005ji}
  Z.~Bern, L.~J.~Dixon and D.~A.~Kosower,
  Phys.\ Rev.\  D {\bf 72} (2005) 125003
  [arXiv:hep-ph/0505055].

\bibitem{Bern:2005hh}
Z.~Bern, N.~E.~J.~Bjerrum-Bohr, D.~C.~Dunbar and H.~Ita,
  JHEP {\bf 0511} (2005) 027
  [arXiv:hep-ph/0507019].

\bibitem{Berger:2006ci}
  C.~F.~Berger, Z.~Bern, L.~J.~Dixon, D.~Forde and D.~A.~Kosower,
  Phys.\ Rev.\  D {\bf 74}, 036009 (2006)
  [arXiv:hep-ph/0604195].

\bibitem{Cachazo:2004zb}
  F.~Cachazo, P.~Svrcek and E.~Witten,
  JHEP {\bf 0410}, 074 (2004)
  [arXiv:hep-th/0406177].

\bibitem{Cachazo:2004by}
  F.~Cachazo, P.~Svrcek and E.~Witten,
  JHEP {\bf 0410}, 077 (2004)
  [arXiv:hep-th/0409245].

\bibitem{Cachazo:2004dr}
  F.~Cachazo,
  arXiv:hep-th/0410077.

\bibitem{Brandhuber:2004yw}
  A.~Brandhuber, B.~J.~Spence and G.~Travaglini,
  Nucl.\ Phys.\ B {\bf 706} (2005) 150
  [arXiv:hep-th/0407214].

\bibitem{Brandhuber:2005kd}
  A.~Brandhuber, B.~Spence and G.~Travaglini,
  JHEP {\bf 0601}, 142 (2006)
  [arXiv:hep-th/0510253].

\bibitem{Xiao:2006vr}
  Z.~Xiao, G.~Yang and C.~J.~Zhu,
  Nucl.\ Phys.\  B {\bf 758} (2006) 1
  [arXiv:hep-ph/0607015].

\bibitem{Bedford:2004nh}
  J.~Bedford, A.~Brandhuber, B.~J.~Spence and G.~Travaglini,
  Nucl.\ Phys.\ B {\bf 712} (2005) 59
  [arXiv:hep-th/0412108].

\bibitem{Britto:2004nc}
R.~Britto, F.~Cachazo and B.~Feng,
  Nucl.\ Phys.\ B {\bf 725}, 275 (2005)
  [arXiv:hep-th/0412103];

\bibitem{Bidder:2005ri}
S.~J.~Bidder, N.~E.~J.~Bjerrum-Bohr, D.~C.~Dunbar and W.~B.~Perkins,
  Phys.\ Lett.\ B {\bf 612}, 75 (2005)
  [arXiv:hep-th/0502028];

\bibitem{Buchbinder:2005wp}
E.~I.~Buchbinder and F.~Cachazo,
  JHEP {\bf 0511}, 036 (2005)
  [arXiv:hep-th/0506126];

\bibitem{Risager:2005ke}
K.~Risager, S.~J.~Bidder and W.~B.~Perkins,
  JHEP {\bf 0510}, 003 (2005)
  [arXiv:hep-th/0507170];

\bibitem{Bern:2004bt}
Z.~Bern, L.~J.~Dixon and D.~A.~Kosower,
  Phys.\ Rev.\ D {\bf 72}, 045014 (2005)
  [arXiv:hep-th/0412210].

\bibitem{Quigley:2004pw}
C.~Quigley and M.~Rozali,
  JHEP {\bf 0501}, 053 (2005)
  [arXiv:hep-th/0410278];

\bibitem{Bedford:2004py}
J.~Bedford, A.~Brandhuber, B.~J.~Spence and G.~Travaglini,
  Nucl.\ Phys.\ B {\bf 706}, 100 (2005)
  [arXiv:hep-th/0410280];

\bibitem{Gorsky:2005sf}
A.~Gorsky and A.~Rosly,
  JHEP {\bf 0601}, 101 (2006)
  [arXiv:hep-th/0510111].

\bibitem{Mansfield:2005yd}
  P.~Mansfield,
  JHEP {\bf 0603} (2006) 037
  [arXiv:hep-th/0511264].

\bibitem{Bergere:1975tr}
  M.~C.~Bergere and Y.~M.~Lam,
  Phys.\ Rev.\  D {\bf 13} (1976) 3247.

\bibitem{Chalmers:1996rq}
  G.~Chalmers and W.~Siegel,
  Phys.\ Rev.\  D {\bf 54} (1996) 7628
  [arXiv:hep-th/9606061].

\bibitem{Itzykson:1980rh}
  C.~Itzykson and J.~B.~Zuber,
  ``Quantum Field Theory,''
{\it  New York, McGraw-Hill (1980).}

\bibitem{Ettle:2006bw}
  J.~H.~Ettle and T.~R.~Morris,
  JHEP {\bf 0608} (2006) 003
  [arXiv:hep-th/0605121].

\bibitem{Feng:2006yy}
  H.~Feng and Y.~t.~Huang,
  arXiv:hep-th/0611164.

\bibitem{Brandhuber:2006bf}
  A.~Brandhuber, B.~Spence and G.~Travaglini,
  arXiv:hep-th/0612007.

\bibitem{Mason:2005kn}
  L.~J.~Mason and D.~Skinner,
  Phys.\ Lett.\  B {\bf 636} (2006) 60
  [arXiv:hep-th/0510262].

\bibitem{Boels:2006ir}
  R.~Boels, L.~Mason and D.~Skinner,
  arXiv:hep-th/0604040.

\bibitem{Boels:2007qn}
  R.~Boels, L.~Mason and D.~Skinner,
  arXiv:hep-th/0702035.
  
\bibitem{Boels:2007gv}
  R.~Boels,
  arXiv:hep-th/0703080.
  
\bibitem{Leibbrandt:1983pj}
  G.~Leibbrandt,
  Phys.\ Rev.\  D {\bf 29} (1984) 1699.

\bibitem{analytics}
 R.~J.~Eden, P.~V.~Landshoff, D.~I.~Olive and J.~C.~Polkinghorne,
 ``The Analytic $S$-Matrix,''
 {\it  Cambridge University Press (1966).}
 
\bibitem{Feynman:1972mt}
  R.~P.~Feynman,
{\it  In *J R Klauder, Magic Without Magic*, San Francisco 1972, 355-375}

\bibitem{Dixon:1996wi}
L.~J.~Dixon,
  lectures published in Boulder TASI 95:539--584
  [arXiv:hep-ph/9601359].

\bibitem{Bern:1993mq}
  Z.~Bern, L.~J.~Dixon and D.~A.~Kosower,
  Phys.\ Rev.\ Lett.\  {\bf 70} (1993) 2677
  [arXiv:hep-ph/9302280].

\bibitem{Bern:1993sx}
Z.~Bern, L.~J.~Dixon and D.~A.~Kosower,
  arXiv:hep-th/9311026.

\bibitem{Bern:1995db}
  Z.~Bern and A.~G.~Morgan,
  Nucl.\ Phys.\  B {\bf 467} (1996) 479
  [arXiv:hep-ph/9511336].

\bibitem{Brandhuber:2007vm}
  A.~Brandhuber, B.~Spence, G.~Travaglini and K.~Zoubos,
  arXiv:0704.0245 [hep-th].

\bibitem{Chakrabarti:2005ny}
  D.~Chakrabarti, J.~Qiu and C.~B.~Thorn,
  Phys.\ Rev.\  D {\bf 72}, 065022 (2005)
  [arXiv:hep-th/0507280].

\end{thebibliography}
\end{document}